\documentclass[12pt]{article}
\usepackage{amsfonts,amsmath} 

\date{{\small Oct. 14, 1999; \ \  rev. Mar. 24, 2000.}}

\input epsf     
\def\be{\begin{equation}}
\def\ee{\end{equation}}
\def\bea{\begin{eqnarray}}
\def\eea{\end{eqnarray}}

\newcommand{\eq}[1]{eq.~(\ref{#1})}

\def\blackbox{{\vrule height 1.3ex width 1.0ex depth -.2ex}
       \hskip 1.5truecm}
\def\Cell{{\mathcal C}}   
\def\c{\mathrm{Const.}}

\def\G  { {\mathcal G} }
\def\T{ {\mathcal T}}     
\def\dist{\operatorname{dist}}   
\def\dist  {{\rm dist}}
\def\E{{\mathbb E}}    

\def\half {{1 \over 2}}

\def\liminf{\mathop{\underline{\rm lim}}}

\def\P{\operatorname{Prob}} 
\def\R{{\mathbb R}}    
\def\C{{\mathbb C}}    
\def\too#1{\parbox[t]{.4in} {$\longrightarrow\\[-9pt] {\scriptstyle #1}$}}
\def\Z{{\mathbb Z}}    

\newcounter{masectionnumber}
\newcommand{\masect}[1]{\setcounter{equation}{0}
\refstepcounter{masectionnumber} \vspace{1truecm plus 1cm} \noindent
    {\large\bf \arabic{masectionnumber}. #1}\par \vspace{.2cm}
      \addcontentsline{toc}{section}{\arabic{masectionnumber}. #1}
    }
\renewcommand{\theequation}
    {\mbox{\arabic{masectionnumber}.\arabic{equation}}}

\newcounter{masubsectionnumber}[masectionnumber]
\newcommand{\masubsect}[1]{
    \refstepcounter{masubsectionnumber} \vspace{.5cm} \noindent
    {\large\em \arabic{masectionnumber}.\alph{masubsectionnumber} #1}
    \par\vspace*{.2truecm}
    \addcontentsline{toc}{subsection}
    {\arabic{masectionnumber}.\alph{masubsectionnumber}\hspace{.1cm}
        #1}
    }

\newtheorem{lem}{Lemma}[masectionnumber]
\newtheorem{thm}[lem]{Theorem}
\newtheorem{prop}[lem]{Proposition}

\newtheorem{df}{Definition}[masectionnumber]

\newenvironment{proof_of}[1]{
    \noindent{\bf Proof of #1:} \hspace*{1em} }{
    \hfill \blackbox\bigskip}
\newenvironment{proof}{\noindent{\bf Proof:}
   \hspace*{1em}}{\hfill \blackbox\bigskip}

\newcommand{\startappendix}{ \setcounter{masectionnumber}{0}
 \renewcommand{\theequation}
    {\mbox{\Alph{masectionnumber}.\arabic{equation}}}
 \renewcommand{\thelem}
    {\mbox{\Alph{masectionnumber}.\arabic{lem}}}
 \renewcommand{\themasectionnumber}
    {\mbox{\Alph{masectionnumber}}}
\addcontentsline{toc}{section}{Appendix }
  }
\newcommand{\maappendix}[1]{          
    \setcounter{equation}{0}
\refstepcounter{masectionnumber} \vspace{1truecm plus 1cm} \noindent
    {\large\bf \Alph{masectionnumber}. #1}\par \vspace{.2cm}

    \addcontentsline{toc}{section}{\Alph{masectionnumber}. #1 }   }

    \newcounter{masubapp}[masectionnumber]
\newcommand{\masubappendix}[1]{        
    \refstepcounter{masubapp} \vspace{.5cm} \noindent
  {\large\em \Alph{masectionnumber}.\alph{masubapp} #1}
    \par\vspace*{.2truecm}

\addcontentsline{toc}{subsection}
 {\Alph{masectionnumber}.\alph{masubapp}\hspace{.1cm} #1 }
      }

\begin{document}
\begin{figure}[t]{  \sf \em {To appear in  \vfill   
 Comm. Math. Phys.} 
 \vfill } \end{figure} 

\title{\vspace*{-.35in}
Finite-Volume Fractional-Moment Criteria for Anderson Localization
                 }

\author{Michael Aizenman $^{a,b}$ \ \ \ \  Jeffrey H. Schenker $^b$ \\
  \vspace*{-0.15truein} \\
  Roland M. Friedrich $^c$ \ \ \ \  Dirk Hundertmark $^{a}$ \\
  \vspace*{-0.05truein} \\
\normalsize \it  Departments of Physics$^{(a)}$ and Mathematics$^{(b)}$,
Princeton University \\
\normalsize \it Princeton, NJ 08544, USA. \\
  \vspace*{-0.2truein}  \\
\normalsize \it ${}^{(c)}$ Student, Theoretische Physik,
ETH-Z\"urich, CH--8093, Switzerland.
      }
\maketitle
\thispagestyle{empty}        

\begin{abstract}
A technically convenient signature of localization,
exhibited by discrete operators with random potentials,
is exponential decay of the fractional
moments of the Green function within the appropriate energy ranges.
Known implications include: spectral localization,
absence of level repulsion, strong form of dynamical localization,
and a related condition which plays a significant role in the
quantization of the Hall conductance in two-dimensional Fermi gases.
We present a family of finite-volume criteria which,
under some mild restrictions on the distribution of the potential,
cover the regime where the fractional moment decay condition holds.
The constructive criteria permit to establish this condition
at spectral band edges, provided there are sufficient
`Lifshitz tail  estimates' on the density of states.
They are also used here to
conclude that the fractional moment condition, and thus
the other manifestations of localization, are
valid throughout the regime covered by the ``multiscale analysis''.
In the converse direction, the analysis rules out fast power-law
decay of the Green functions at mobility edges.
\end{abstract}

\vskip .25truecm
\newpage

  \vskip .25truecm         
    \begin{minipage}[t]{\textwidth}
    \tableofcontents
    \end{minipage}
   \vskip .5truecm
\newpage

\masect{Introduction}
\label{sect:intro}
\vskip-0.6cm

\masubsect{Overview}

Operators with extensive disorder are known to have spectral
regimes (energy ranges) where the spectrum consists of a dense
collection of eigenvalues corresponding to exponentially localized
eigenfunctions. This phenomenon is of relevance  in different
contexts; e.g., it plays a role in the
conductive properties of metals \cite{An,MT,MS},
in  the quantization of
Hall conductance~\cite{Halp82,NTW,AS2,BES,AG}, and in the
emerging subject of optical crystals~\cite{FiKl-SIAM}.

Most of the mathematical results on localization for
operators with random potential in dimensions $d>1$  have been
derived using the {\em multiscale analysis} introducted by
Fr\"ohlich and Spencer~\cite{FS} (and later evolved through
various other works).   For discrete systems
there is an alternative approach,
based on the analysis of the Green function's
{\em fractional moments}~\cite{AM}.  This approach has so far
been developped for only a subset of the localization
regime, but were it applies it yields  somewhat stronger
conclusions (through elementary arguments).  In this work
we present a further extension of that method.
In particular, we derive a  family of constructive
finite-volume criteria for the exponential decay for the
fractional moments of Green functions.  This decay condition
is a technically convenient characterization of
localization, for it is known to imply spectral localization,
absence of level repulsion, dynamical localization
(in a strong exponential sense) and a
related condition which plays a significant role in the
quantization of the Hall conductance in two-dimensional Fermi gases.
The constructive criteria are used to prove that
for the discrete random operators described below
all these properties
hold throughout the regime of localization -- if that is defined
through either the criteria of the multiscale analysis or
those presented here.   The constructive criteria
also preclude fast power-law decay of the Green
functions at mobility edges.

A guiding example for the operators discussed here
is the discrete Schr\"odinger operator, acting in
 $\ell^2(\Z^d)$:
\begin{equation}
H_{\omega} \ = \ T + \lambda V_{\omega} \; ,
\label{eq:proto}
\end{equation}
with $T$ denoting the off-diagonal part, whose matrix elements are
referred to as the {\em hopping terms}, and $V_{\omega}$ a random
multiplication operator -- referred to as the {\em potential}.
The symbol  $\omega$ represents a particular
realization of the disorder, in this case the potential
variables $\{V_{\omega}(x)\}$, and
$\lambda$ serves as the disorder strength parameter.

For the discrete Schr\"odinger operator
\begin{equation}
T_{u,v} \ = \ \begin{cases}
    1 & \mbox{ if }|u-v| = 1 \; ,\\
    0 & \mbox{ if }|u-v| \neq 1 \; ,
\end{cases}
\label{eq:lap}
\end{equation}
and the random potential is given by a collection of independent
identically distributed random variables, $\{V_{\omega}(x)\}_{x\in
\Z^d}$.  However, we shall also consider a more general class of
operators, allowing the incorporation of magnetic fields, periodic
terms, and off-diagonal disorder (see Section~\ref{sect:gen}). We
focus on the case of extensive disorder, where the distribution of
the random operator $H_{\omega}$  is either translation invariant,
or at least gauge equivalent to shifts by multiples of basic
periods (i.e. invariant under periodic magnetic shifts).

Our main goal is to present a sequence of
finite-volume criteria for localization, which permit to conclude
that the following fractional-moment condition is satisfied
in some energy interval $[a,b]\in \R$:
\begin{equation}
\E(|<x| {1 \over H_{\omega}-E - i \eta } | y>|^s) \ \le \ A(s) \
e^{-\mu(s)
|x-y|}  \ , \label{eq:fm}
\end{equation}
for all $E\in [a,b]$, $\eta \in \R$,  and suitable $s\in (0,1)$.
$\E(\cdot )$ represents here the average over the disorder, {\it
i.e.} the random potential.

Needless to say, the bound (\ref{eq:fm}) is of interest mainly in
situations  where the energy $E$ is within the spectrum, {\it
i.e.} $[H_{\omega}-E]^{-1}$ is an unbounded operator
  and the  exponential
 decay occurs only due to the localization of the eigenvalues with
 energies within the interval $[a,b]$.
As in ref.~\cite{AM}, fractional powers are used in
order to avoid infinity, however the value of $ 0< s < 1$
at which \eq{eq:fm} is derived is of almost no importance
(if \eq{eq:fm} holds for a particular value of $s$,
then it will hold for all $s < \tau$, where $\tau < 1$ is
a number which depends only on the regularity of the probability
distribution of $V_\omega(x)$, see Appendix -- Lemma~\ref{lem:allforone}).

For the systems considered here, \eq{eq:fm} is known to imply
various other properties, mentioned above, which are commonly
associated with localization.  More explicitely:
\begin{itemize}
\item[i.]  {\it Spectral localization (\cite{AM} - using \cite{SiWo}):}
The spectrum of $H_{\omega}$
within the interval $(a,b)$ is almost-surely of the pure-point type,
and the corresponding eigenfunctions are exponentially localized.
\item[ii.] {\it  Dynamical localization (\cite{Ai94}, expanded here
in Appendix \ref{sect:dynamical}):}
wave packets with energies in the
specified range do not spread --
\begin{equation}
\E\left( \sup_{t\in \R} |<x| \ e^{-itH} P_{H\in [a,b]}\ |y>|
\right ) \ \le \tilde A e^{-\tilde \mu |x-y|} \label{eq:dyn}
\end{equation}
\item[iii.] {\it Exponential decay of the projection kernel
(\cite{AG})}; the condition  expressed in a bound similar to
\eq{eq:dyn} for $\E( |<x|\ P_{H\le E}\ |y>|)$ with  $E\in [a,b]$.
This condition plays an important role in the quantization of Hall
conductance, in the ground state of the two dimensional electron
gas with Fermi level $E_{F}\in [a,b]$ \cite{BES,AS2,AG}.
\item[iv.] {\it Absence of level repulsion (\cite{Min}).}
Minami  has shown that \eq{eq:fm} implies, for operators
of the type considered here, that in the
range $[a,b]$ the energy gaps have Poisson-type statistics.
\end{itemize}
The fractional moment condition has already been established for
certain regimes: extreme energies, as well as all energies at high
enough disorder \cite{AM}, and also for weak disorder but far
enough from the unperturbed spectrum \cite{Ai94}. The results
presented below permit to extend it to  band edges, provided there
are sufficient
 `Lifshitz tail  estimates' on the density of states
(ref.~\cite{FigPast,BCH,KSS,Stollmann,Klopp99}), and to other
regimes mapped by a sequence of constructive criteria.


\masubsect{The finite-volume criteria}
\label{sect:main}

Our main results admit a number of variations.  In this section we
present a formulation which is natural for the prototypical
example of the discrete random Schr\"odinger operators, {\it i.e.}
Hamiltonians of the form (\ref{eq:proto}) with $T$ the discrete
Laplacian (given by (\ref{eq:lap})). In Section~\ref{sect:gen} we
formulate various extensions of the results, including to
operators incorporating magnetic fields and to operators with
hopping terms of unbounded range.

The results are derived under some mild regularity assumptions on
the probability distribution of the variables $\{
V_{\omega}(x)\}_{x\in \Z^d}$ which form the random potential. For
simplicity we address ourselves here to the {\em IID} case: the
potential variables are independent with a common probability
distribution $\rho(dV)$.   The assumption is then that $\rho(dV)$
satisfies the regularity conditions listed below, $R_1(s)$ or
$R_2(s)$. However, the independence is not essential.  What
matters is that  the stated regularity condition be satisfied,
with a uniform constant,  by the  conditional distribution of each
of the potential variables, conditioned on arbitrary values of the
other potentials.

The two regularity conditions mentioned here are:
\begin{description}
  \item[$R_1(s)$:] A probability distribution $\rho(dV)$, on $\R$, is said
  to be {\em $s$-regular}, or to satisfy the condition $R_1(s)$ at
  some $0<s\le 1$,  if there exists $C < \infty$ such that
  \begin{equation}
  \label{eq:rregular}
  \rho(a - \epsilon, a + \epsilon) \leq C \epsilon^s.
  \end{equation}
  \item[$R_2(s)$:] The probability distribution $\rho(dV)$ is said
  to have the {\em decoupling property}  $R_2(s)$, with
  some $0<s\le 1$,  if there exists $C < \infty$ such that
  for any pair of functions  $f$ and $g$ of the form
  \begin{equation}
  f(V) =  {1 \over V - a}, \qquad
  g(V)  =  {V - b \over V - c} \; ,
  \end{equation}
  with $a,b,c \in {\mathbb C}$,
  the expectation of the product can be dominated as follows:
  \begin{equation}
  \E \left (|f(V)|^s |g(V)|^s \right ) \ \leq \ C \ \E \left ( |f(V)|^s
  \right ) \ \E \left ( |g(V)|^s \right ),
  \label{eq:decoupling}
  \end{equation}
  The smallest $C$ such that \eq{eq:decoupling} holds for all
  $a,b,c \in {\mathbb C}$ is called
  here the {\em decoupling constant} for $\rho$, and is denoted by
   $D_s(\rho)$.
\end{description}
A sufficient condition for $R_2(s)$ is
  that $\rho$ have bounded support and satisfy
  $R_1(\tau)$ for some $\tau > 4s$ (see Appendix
  \ref{sect:decoupling}; related discussion is found in
  Refs.~\cite{AM,AG}.)

In Appendix \ref{sect:moment} we show that given any
$\tau$-regular measure $\rho$ and any $s < \tau$, there is a
finite constant $C$ such that for any $2 \times 2$ self adjoint
matrix $A_{2\times 2}$
\begin{equation}
\int \int \rho(du) \rho(dv)  \left | \left [ \left ( A_{2 \times
2} +
\begin{pmatrix}
  u & 0 \\
  0 & v
\end{pmatrix}
\right )^{-1} \right ]_{i,j}\right |^s \ \le \ C \; ,
\label{eq:2by2average}
\end{equation}
where $[ \cdot ]_{i,j}$ denotes the $i,j$ matrix element with $i,j
= 1,2\, $.  Throughout this work, we denote by $C_s$ the smallest
value of $C$ at which
(\ref{eq:2by2average}) holds.  For $\rho(dV)$ which also satisfy
 $R_2(s)$ we let:
$\widetilde C_s = C_s \cdot D_s(\rho)^2$.

For $\Lambda \subset \Z^d$ we denote by
$H_{\Lambda ; \omega}$ the operator obtained from
$H_{\omega}$ by ``turning off'' the hopping terms outside
$\Lambda$.
Thus, the restriction of $H_{\Lambda ; \omega}$ to
$\ell^2 (\Lambda)$ (considered as a subspace of $\ell^2(\Z^d)$),
is nothing but
$H_\omega$ with the Dirichlet boundary conditions on the boundary
of $\Lambda$.

We also denote  by $\Gamma(\Lambda)$ the set of the
nearest-neighbor bonds reaching out of $\Lambda$ ({\it i.e.} pairs
with one site in $\Lambda$ and the other outside), by
$\Lambda^{+}$ the collection of sites within distance $1$ from
$\Lambda$, and by $|\Gamma(\Lambda^+)|$ the number of bonds
reaching out of that set.  These notions will be generalized in
Section~2.a.

Following are  our basic results for operators of the form
(\ref{eq:proto}).

\begin{thm}
\label{thm:1} Let $H_{\omega}$ be a random Schr\"odinger operator
with the probability distribution of the potential $V(x)$
satisfying the regularity condition $R_1(\tau)$ and fix $s <
\tau$.
If for some $z\in \C$ (possibly real) and some finite region
 $ \Lambda\subset \Z^d $
which contains the origin $0$:
\begin{equation}
b(\Lambda, z) \ := \ \sup_{W\subset \Lambda }
 \ \left\{ |\Gamma(\Lambda^+)| \ {C_s \over \lambda^s} \ \sum_{<u,u'> \in
\Gamma(\Lambda) } \E\left(|<0| {1 \over H_{W ; \omega} - z} |u>|^s
\right) \right\}  \  < \ 1 \; , \label{eq:cond1}
\end{equation}
then there are some $\mu(s) > 0$ and
$A(s) < \infty $ --- which depend on
the energy $z$ only through the bound $b(\Lambda,z)$ --- such
that for any region $\Omega \subset \Z^d$
\begin{equation}
\label{eq:thm1}
 \E_{\pm i 0}\left( |<x| {1 \over  H_{\Omega;\omega} - z }
|y> |^s \right)\ \le \ A(s) \  e^{-\mu(s) \, |x-y| } \; .
\end{equation}
\end{thm}

The subscript of  $\E_{\pm i 0} $, in (\ref{eq:thm1}) is to
be interpreted as saying  that the bound is
valid for either of the two limiting expressions:
\begin{equation}
\lim_{\eta \searrow 0} \E\left( |<x| {1 \over H_{\Omega;\omega}
- E -\!(+)\, i \eta } |y> |^s \right) \; .
\end{equation}
The ``cutoff'' $\pm i\eta$ is needed for an
unambiguous interpretation in case $z$ is a real energy
($ E $)  within the spectrum of $H$.
For the random operators considered here it is well understood
that: i) the expectation may be exchanged with the limit $\eta
\searrow 0$, ii) it suffices to verify the uniform bounds
(\ref{eq:thm1}) for finite regions, and iii) the finite volume
expectations are continuous in $\eta$.
In the proofs we shall be dealing with finite systems;
the subscript will, therefore,  be omitted there.

Let us note that already the special case $\Lambda = \{ 0\}$ is of
interest.  It provides the following variant of the single-site
criterion of ref.~\cite{AM} (which is, in fact, a bit simpler
since it does not invoke the {\em decoupling lemma}).

\noindent{\bf Corollary} {\it For the random Schr\"odinger
operator a sufficient condition for localization (\ref{eq:fm}) is
that for all $E\in [a,b]$
\begin{equation}
2d (2d -1) \ {C_s \over \lambda^s}  \ \int {1 \over |\lambda V -
E|^s} \ \rho(dV) \ < \ 1 \; .
\label{eq:singlesite}
\end{equation}
 }

Just as the main result of ref.~\cite{AM}, the above criterion
permits to easily conclude localization for the cases of high
disorder or extreme energies.  However, we may now  move beyond
that. By testing the hypothesis of Theorem~\ref{thm:1} in the
increasing sequence of volumes $\Lambda = [-L,L]^d$, one may
extend the conclusion to increasing regimes in the `energy
$\times$ disorder plane'.  In fact, it is easy to see that for
each energy at which the strong localization condition
(\ref{eq:thm1}) is satisfied, the hypothesis (\ref{eq:cond1}) will
be met at all sufficiently large $L$.  (This may, however,  be far
from a practical test, as the necessary computation may be rather
difficult for large $L$).

Observant readers may note that the conclusion of
Theorem~\ref{thm:1} provides not only the localization condition
\eq{eq:fm}, but it also rules out {\em extended boundary states}.
The flip side of this observation is that if such states are
present in some geometry, {\it e.g.} the half space, then the
hypothesis of Theorem~\ref{thm:1} will fail to be satisfied even
if the operator exhibits localization in the bulk.   Therefore, we
present also the following result which  permits to establish bulk
localization regardless of the possible presence of extended
boundary states.

\begin{thm}
\label{thm:2} Let $H_{\omega}$ be a random Schr\"odinger operator
with the probability distribution of the potential $V(x)$
satisfying $R_1(\tau)$ and $R_2(s)$, for some $s < \tau$. If for
some $z \in \C$ and some finite region $0 \in \Lambda \subset
\Z^d$
\begin{equation}
 \left ( 1 + {\widetilde C_s \over \lambda^s} \ |\Gamma(\Lambda)| \right )^2
   \sum_{<u,u'> \in \Gamma(\Lambda)}
            \E \left (
                |<0| {1 \over
                    H_{\Lambda;\omega} - z} |u>|^s
            \right ) \ < \ 1 ,
\label{eq:cond2}
\end{equation}
then $H_{\omega}$ satisfies the
fractional-moment condition (\ref{eq:fm}),
and there exist $\mu(s) > 0, A(s) < \infty $ so that for any
region $\Omega \subset
\Z^d$,
\begin{equation}
\E_{\pm i 0}\left( |<x| {1 \over  H_{\Omega;\omega}-z} |y> |^s \right)\ \le
\ A(s)\  e^{-\mu(s)\,  \dist_{\Omega}(x,y)} \; ,
\label{eq:thm2}
\end{equation}
with
\begin{equation}
\dist_{\Omega}(x,y)= \min\{|x-y|, [\dist(x,\partial \Omega)
+ \dist(y,\partial \Omega)] \} \; .
\label{eq:dist}
\end{equation}
\end{thm}

Let us add that,
as in Theorem~\ref{thm:1}, $A(s)$ and $\mu(s)$ of (\ref{eq:thm2})
depend on $z$ only  through the value of the LHS in \eq{eq:cond2}.

The modified metric, $\dist_{\Omega}(x,y)$,  is a distance
function relative to which the entire boundary of $\Omega$ is
regarded as one point. It permits us to state that there is
exponential decay in the bulk without ruling out non-exponential
decay along the boundary. We supplement the last result by the
following observation.

\begin{thm}
Let $H_{\omega}$ be a random operator given by \eq{eq:proto}, with
the probability distribution of the potential $V(x)$ satisfying
$R_1(\tau)$ and $R_2(s)$, for some $s < \tau$. If at some energy
$E$ (or $z\in \C$) the localization condition  (\ref{eq:fm}) is
satisfied, with
some $A< \infty$ and $\mu> 0$, then for all  large enough
(but finite) $L $
the condition (\ref{eq:cond2}) is met for $\Lambda = [-L,L]^d$.
\label{thm:3}
\end{thm}

The statement is a bit less immediate than the analogous
claim for Theorem~\ref{thm:1}.  We shall therefore include the
proof below.

It is natural to compare the above criteria for localization
with those of the multiscale analysis.  The two methods share the
basic
feature that the analysis requires an initial condition which one
may expect to be met in a finite system provided  its linear size
is  of the order of the localization length, or larger. However,
for the method presented here if a suitable input is received on
some scale, then the analysis can proceed using steps, or blocks,
of only that size.  An important difference in the results is that
the fractional moment condition yields exponential decay for the
\underline{expectation values}, which  are important for some of
the conclusions listed above.  Such bounds have not been derived
by methods based on
the multiscale analysis, since (at least without further improvement)
the  bounds the latter yields on the ``error terms'', i.e., the
probabilities of ``bad blocks'',  decay not faster than
 $\exp[- (\log L / \log L_o)^{\alpha}]$.  This rate  is
 faster than any power of $L$, but in itself not
fast enough to imply exponential bounds for the mean values.
However, it should be noted that the extension of the present
method to
operators in the continuum, for which a number of basic
localization results have been established using the multiscale
analysis \cite{CH,FiKl,KSS},  is still unaccomplished. Also
not covered are discrete operators with the
potential assuming discrete values (e.g., $V_{\omega}(x)= \pm 1$
\cite{CKM}).

In  Section~\ref{sect:implications} we discuss various
implications of the basic results.  In particular it is shown
that, for discrete random operators of the type considered here,
the fractional moment condition (\ref{eq:fm}) is satisfied
throughout the regime in which the multiscale analysis applies
(see Theorem~\ref{thm:multiscale}). This carries the further
implication that the properties listed above hold throughout the
entire regime for which localization can be proven by any of the
known methods.  One of those properties is a strong form of
dynamical localization, on which more is said in
Appendix~\ref{sect:dynamical}.


\masect{Proofs of the main results} \label{sect:proof}

\masubsect{Some useful notation}

The proofs of the above statements will be presented in terms
which permit a direct extension to operators with more general
hopping terms.  We start by generalizing the notation; in
particular, the sets $\Lambda^{+}$ and $ \Gamma(\Lambda)$ will
be made to depend implicitly on the operator $T$.

In the study of  $H_{\Omega;\omega}$ we shall often consider
`depleted' Hamiltonians, $ H_{\Omega;\omega}^{(\Gamma)} $, obtained by setting
to zero the operator's non-diagonal matrix elements ({\em hopping terms})
along some collection of ordered pairs of sites (referred to here
as {\em bonds})  $\Gamma \subset \Z^d \times \Z^d$.
The difference is the operator $T^{(\Gamma)}$, with
\begin{equation}
T^{(\Gamma)}_{x,y} = \begin{cases}
        T_{x,y} \ & \mbox{ if } <x,y> \in \Gamma \mbox{ or } <y,x> \in
        \Gamma
        \\
        0 \ & \mbox{ if } <x,y> \not \in \Gamma \mbox{ and } <y,x> \not
        \in \Gamma   \; ,
        \end{cases}
\end{equation}
so that
\begin{equation}
H_{\Omega;\omega} \ = \ H_{\Omega;\omega}^{(\Gamma)} + T^{(\Gamma)}
\; .
\end{equation}

Typically,  $\Gamma$ will be a collection of bonds which forms
the `cut set' of some $W\subset \Z^d$, i.e.,
the set of bonds with $T_{x,y}\neq 0$ connecting sites in $W$ with
sites in its complement.  Thus we denote
\begin{equation}
\Gamma(W) \ = \ \left\{ <u,u'> | \ u \in
          W, u'\in \Z^d
          \backslash W, \mbox{ and }
          T_{u,u'}\neq 0    \right\} \; ,
\end{equation}
and also
\begin{equation}
      W^+ \ = \ W \cup \left \{ u' \in \Z^d |
        \ T_{u,u'} \neq 0
        \mbox{ for some } u \in W \right \} \; .
\end{equation}
The number of elements ({\it i.e.} bonds) in $\Gamma$ is denoted
$|\Gamma|$.

In addition, we use the ``Green function'' notation:
\begin{equation}
G_{\Omega;\omega}(x,y;z) = <x| {1 \over H_{\Omega;\omega} - z} |y>
\; ,
\end{equation}
with $G_{\Omega;\omega}^{(\Gamma)}(x,y;z)$ defined
correspondingly. Often, where it is obvious from context that an
operator is a random variable, we shall suppress the subscript
$\omega$.

In broad terms, the strategy for the proof is to derive a bound on
the average Green function, of the form
\begin{equation}
\E \left ( |G_\Omega(x,y;z)|^s \right ) \ \le  \ \sum_{< u ,u' >
\in \Gamma(\Lambda(x))} \gamma_{\Lambda(x)}(< u ,u' >)|T_{u,u'}|^s \
\E \left ( |G_{\Omega}^{(\Gamma(\Lambda(x))}(u',y;z)|^s \right ) \; ,
\label{eq:protobound}
\end{equation}
for all $y \in \Z^d \backslash \Lambda(x) $, where:
$\Lambda(x)= \{ x+y : y\in \Lambda \}$ is a finite neighborhood of $x$,
 translate of some fixed region $\Lambda \ni 0$,
and $\gamma_{\Lambda(x)}$ is a quantity which is small when the
typical values of the finite volume Green function between $x$ and
the boundary of $\Lambda(x)$ are small (in a suitable sense).

An inequality of the form (\ref{eq:protobound}) is particularly
useful when
\begin{equation}
\sum_{<u,u'> \in \Gamma(\Lambda(x))} \gamma_{\Lambda(x)}(<u,u'>)
\ |T_{u,u'}|^s
 \ < \ 1 \; ,
\end{equation}
since in that case \eq{eq:protobound} is akin to the statement
that $\E \left ( |G_\Omega(x,y;z)|^s \right )$ is a strictly
subharmonic function of $x$, as long as $|x-y| > {\rm diam} |\Lambda|$,
and thus --- if it is also  uniformly
bounded (which it is) --- it decays exponentially.

The first step towards a bound of the form (\ref{eq:protobound})
is, naturally, the resolvent identity:
\begin{eqnarray}
\nonumber G_{\Omega,\omega} \ & = & \
 G_{\Omega,\omega}^{(\Gamma)} -
   G_{\Omega,\omega}^{(\Gamma)} \cdot T^{(\Gamma)} \cdot
      G_{\Omega,\omega}
        \\
       & & \nonumber \\
       & = & \ G_{\Omega,\omega}^{(\Gamma)} -
  G_{\Omega,\omega} \cdot  T^{(\Gamma)} \cdot
   G_{\Omega,\omega}^{(\Gamma)}
\label{eq:firstorder}
\end{eqnarray}
(written here in the operator form). However, one then reaches an
obstacle, since the quantity whose mean needs to be estimated is a
product of two Green functions which are not independent.
For some time now this co-dependence has been the main obstacle on
the road to an argument along the lines outlined above, since
otherwise the general strategy applied here is well familiar from
its various successful applications in the context of the
statistical mechanics of homogeneous systems
(\cite{Ham,DoSh,Simon,Lieb,AiNew}), and the other auxiliary tools
specific to the present context have in essence been available
since ref.~\cite{AM}.
The co-dependence problem is solved here through a
second application of the resolvent identity (followed by a
decoupling argument of a familiar type).
In fact, a similar tactic was applied by
von Dreifus to the mean correlation functions, in a study of the
phase transitions in disordered
ferromagnetic models  \cite{vD91} (as we learned from T. Spencer
after the completion of the first draft of this work).

The two applications of the resolvent identity,
for which the depletion sets $\Gamma_1$ and $\Gamma_2$
need not coincide,  may be combined by
starting our argument from the identity:
\begin{equation}
G_{\Omega} \ = \
 G_{\Omega}^{(\Gamma_1)} -
   G_{\Omega }^{(\Gamma_1)} \cdot T^{(\Gamma_1)} \cdot
      G_{\Omega}^{(\Gamma_2)}
      + G_{\Omega }^{(\Gamma_1)} \cdot T^{(\Gamma_1)} \cdot
      G_{\Omega } \cdot T^{(\Gamma_2)} \cdot
      G_{\Omega}^{(\Gamma_2)}  \; \;   .
\label{eq:secondorder}
\end{equation}
Readers familiar with the current techniques may note that once
the middle term $G_\Omega$ is replaced by a uniform bound, the
remaining expression can be made free from co-dependence by an
appropriate choice of $\Gamma_1$ and $\Gamma_2$. The rest are
technicalities, to which we turn next.

\masubsect{Key Lemmas} \label{sect:lemmas}

We shall now present three Lemmas which will be used in the
proofs of our main results.
The first is a known estimate which provides
the afore-mentioned uniform upper bound.

\begin{lem} \label{lem:1}
Let $V(x)$ be a random potential satisfying the regularity
condition $R_1(\tau)$. Then for each $s< \tau$, any region
$\Omega$,  and any random operator of the form (\ref{eq:proto})
\begin{equation}
\E \left ( |G_\Omega(x,y;z)|^s \right )\  \leq \ {C_s \over
\lambda^s} \; , \label{eq:apriori}
\end{equation}
for all $z \in \C$.
\end{lem}
The statement is an immediate consequence of a version of the Wegner
estimate which we present in the appendix. (See
lemma~\ref{lem:fracmom}; also \eq{eq:conditional} below.)

Next is our new bound.

\begin{lem}
Let $H_{\omega}$ be a random operator given by \eq{eq:proto} with
the probability distribution of the potential $V(x)$ satisfying
the regularity condition $R_1(\tau)$, and let $W$ be a subset of
$\Omega$.  Then, denoting $\widetilde \Gamma = \Gamma(W^{+})$ and
$\Gamma = \Gamma(W)$,
 for all $z \in \C$:
\begin{enumerate}
\item
The following `depleted-resolvent bound' holds for any pair of
sites $x \in W$,  $y \in \Omega \backslash W^+$,
\begin{equation}
\E \left ( |G_\Omega(x,y;z)|^s \right ) \
    \leq \  \gamma(W) \sum_ {<v,v'> \in \widetilde \Gamma} |T_{v,v'}|^s \
         \E \left (| G_{\Omega \backslash W^+}(v',y;z)|^s \right ) \; ,
\label{eq:d-r}
\end{equation}
with
\begin{equation}
\gamma(W) \ = \  {C_s \over \lambda^s}
         \sum_{<u,u'> \in \Gamma} |T_{u,u'}|^s \
         \E \left ( | G_{W}(x,u;z)|^s \right
         ) \; .
\end{equation}
\item If, furthermore, the probability distribution of the
potential satisfies also $R_2(s)$ then the following bound holds
for any pair of sites $x \in W$, $y \in \Omega \backslash W$,
\begin{equation}
\E \left ( |G_\Omega(x,y;z)|^s \right ) \
    \leq \   \sum_ {<v,v'> \in \Gamma} \gamma_x(<v,v'>) \ |T_{v,v'}|^s \
         \E \left (| G_{\Omega \backslash W}(v',y;z)|^s \right ) \; ,
\label{eq:d-r2}
\end{equation}
with
\begin{multline}
\gamma_x(<v',v>) \ = \
    \E \left ( | G_W(x,v';z)|^s \right)
    \\ + \ {\widetilde C_s \over \lambda^s} \sum_{<u,u'> \in \Gamma }
    |T_{u,u'}|^s
    \E \left ( | G_W(x,u;z)|^s \right)    \; .
\label{eq:gamma}
\end{multline}
\end{enumerate}
\label{lem:diagram}
\end{lem}

\begin{figure}[htb]
    \begin{center}
    \leavevmode
        \epsfxsize=4in
  \epsfbox{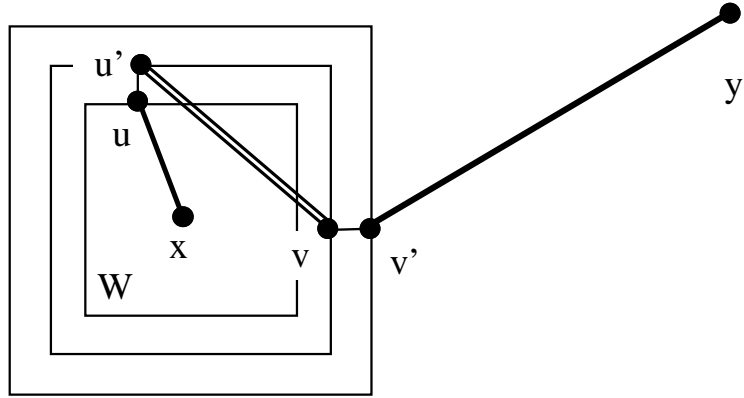}
\caption{\footnotesize
\underline{Diagramatic depiction of the bound
(\ref{eq:diagramatic}) on $G(x,y; z)$, for $x,y \in \Z^d$ and $z\in \C$.}
The long solid lines are `depleted Green functions', the two short
segments correspond to the hoping terms ($T$) and the double line
is a full Green function.  Once the latter is replaced by a uniform
upper bound, the expectation value of the product of the remaining terms
factorizes. }
\label{fig:beads}
\end{center}
\end{figure}

\begin{proof}
Both results follow from the second-order resolvent identity
\eq{eq:secondorder}, which yields:
\begin{multline}
G_{\Omega}(x,y;z)
 \ =  \ G_{\Omega}^{(\Gamma_1)}(x,y;z) \ - \
 <x|  G_{\Omega}^{(\Gamma_1)} \ T_{\Omega}^{(\Gamma_1)} \
 G_{\Omega}^{(\Gamma_2)} |y> \\ + <x|
G_{\Omega}^{(\Gamma_1)} \ T_{\Omega}^{(\Gamma_1)} \ G_{\Omega} \
T_{\Omega}^{(\Gamma_2)} \  G_{\Omega}^{(\Gamma_2)} |y> \; .
\label{eq:secondordermatelts}
\end{multline}

For the proof of the first claim, we take $\Gamma_1 = \Gamma =
\Gamma(W)$ and $\Gamma_2 = \widetilde \Gamma = \Gamma(W^+)$. Then,
the first term of \eq{eq:secondordermatelts} is zero because
$\Gamma(W)$ decouples $x$ and $y$ and the second term is zero
because $\Gamma(W^+)$ decouples $W^+$ and $y$. Thus
\begin{equation}
G_{\Omega}(x,y;z)
        \ = \ \sum_{ \substack{
        <u,u'> \in \Gamma \\
        <v,v'> \in \widetilde \Gamma}}
    T_{u,u'} \ T_{v,v'} \
    G_{\Omega}^{(\Gamma)}(x,u;z)
    G_\Omega(u',v;z)
    G_{\Omega}^{(\widetilde \Gamma)}(v',y;z) \; \; .
\label{eq:diagramatic}
\end{equation}
It follows that for any $ s \in (0,1) $
\begin{multline}
\E \left ( |G_{\Omega}(x,y;z)|^s
        \right )  \\ \leq
    \sum_{ \substack{
        <u,u'> \in \Gamma \\
        <v,v'> \in \widetilde \Gamma}}
        |T_{u,u'}|^s |T_{v,v'}|^s \
    \E \left (
        |  G_{\Omega}^{(\Gamma)}(x,u;z)
    G_\Omega(u',v;z)
      G_{\Omega}^{(\widetilde \Gamma)}(v',y;z)|^s
     \right ) \; .
\label{eq:halffirstdiagram}
\end{multline}
(note that for $0 < s < 1$: $|a+b|^s \leq |a|^s + |b|^s$.)

In estimating the terms on the right hand side of
\eq{eq:halffirstdiagram} let us consider first the conditional
expectation of the central factors, $G_\Omega(u',v;z)$.  Only
these factors depend on the values of the potential at $u'$ and
$v$,
and therefore they can
be replaced by their conditional expectation
$         \E \left ( \left .
           |G_\Omega(u',v;z)|^s
            \right | \{V(q)\}_{q \in \Omega \backslash \{u',v\}}
    \right ) $.
As will be proven in the appendix, under the regularity condition
$R_1(\tau)$ these are uniformly bounded (Lemma~\ref{lem:fracmom}):
\begin{equation}
 \E \left ( \left .
        |G_\Omega(u',v;z)|^s \right |
        \{V(q)\}_{q \in \Omega \backslash \{u',v\}} \right )
    \ \leq  \ {C_s \over \lambda^s} \; .
\label{eq:conditional}
\end{equation}
(The proof involves a reduction to a two-dimensional problem via
the Krein formula, and a two-dimensional Wegner-type estimate.)

Once the central factor in each expectation on the right hand side
of \eq{eq:halffirstdiagram} is replaced by the above bound, what
remains there are  two independent random variables which are ${|
G_{\Omega}^{(\Gamma)}(x,u;z)|^s = |G_W(x,u;z)|^s }$ and ${|
G_{\Omega}^{(\widetilde \Gamma)}(v',y;z)|^s = | G_{\Omega
\backslash W^+ }(v',y;z)|^s}$ . The expectation now factorizes, and
the resulting expression yields the first claim of the Lemma.

For the second claim, we take $\Gamma_1 = \Gamma_2 = \Gamma =
\Gamma(W)$.  Once again the first term of
\eq{eq:secondordermatelts} is zero because $\Gamma(W)$ decouples
$x$ and $y$.  However, the second term is non-zero, and we obtain
\begin{multline}
\E \left ( |G_{\Omega}(x,y;z)|^s
        \right )  \\
\begin{split}
        \leq \ & \sum_{ <v,v'> \in \Gamma}
        |T_{v',v}|^s \E \left (
        | G_{\Omega}^{(\Gamma)}(x,v;z)
         G_{\Omega}^{(\Gamma)}(v',y;z)|^s
        \right ) \\ &  \quad + \
    \sum_{ \substack{
        <u,u'> \in \Gamma \\
        <v,v'> \in \Gamma}}
        |T_{u,u'}|^s |T_{v,v'}|^s \
    \E \left (
        | G_{\Omega}^{(\Gamma)}(x,u;z)
    G_\Omega(u',v;z)
     G_{\Omega}^{(\Gamma)}(v',y;z)|^s
     \right ) \; .
\end{split}
\end{multline}
At this point we may not use the previous argument, since in the
last expectation $V(v)$ affects each of the first two factors and
$V(u')$ affects each of the last two factors.  However, the
dependence of each of these factors on the potentials is of a
particularly simple form: they are ratios of two functions
(determinants) which are separately linear in each potential
variable. Using the decoupling hypotheses, {\it i.e.} the
regularity conditions $R_1(\tau)$  and $R_2(s)$, the expectation
may be bounded by the product of expectations. Specifically, we
prove  in Lemma~\ref{lem:decouplinginequalities} that:
\begin{multline}
\E \left (
        | G_{\Omega}^{(\Gamma)}(x,u;z)
    G_\Omega(u',v;z)
     G_{\Omega}^{(\Gamma)}(v',y;z)|^s
     \right )
    \\ \le \ {\widetilde C_s \over \lambda^s} \E \left (
        | G_{\Omega}^{(\Gamma)}(x,u;z)
         G_{\Omega}^{(\Gamma)}(v',y;z)|^s
        \right ) \; .
\end{multline}

Once again, we are left with a product of two independent random
variables, $| G_{\Omega}^{(\Gamma)}(x,u;z)|^s = | G_W(x,u;z)|^s$
and $| G_{\Omega}^{(\Gamma)}(v',y;z)|^s = | G_{\Omega \backslash W
}(v',y;z)|^s$.  The factorization of the remaining expectation
yields the second claim of the Lemma, \eq{eq:d-r2}.
\end{proof}

The above Lemma provides a bound for the Green function in terms
of its depleted versions.  This suffices for the derivation of the
first of our two main Theorems (Thm~\ref{thm:1}). However, this
does not suffice for the second Theorem, Thm~\ref{thm:2}, for
which we shall use an inequality that is linear in the original
function.  That ``closure'' will be attained with the help of the
following bound on the depleted resolvent in terms of the full
one.

\begin{lem}
Let $H_{\Omega,\omega}$ be a random operator in $\ell^2(\Omega)$,
 $\Omega \subseteq Z^d$, given by \eq{eq:proto}, with
the probability distribution of the potential $V(x)$ satisfying
the regularity conditions $R_1(\tau)$ and $R_2(s)$ for some $s <
\tau$.    Let $W$ be a subset of $\Omega$. Then, the following
holds for any pair of sites
$u,y \in \Omega \backslash  W$, and every $z \in \C$
\begin{multline}
 \E \left ( | G_{\Omega \backslash W}(u,y;z)|^s \right ) \
        \le \ \E \left ( |G_{\Omega}(u,y;z)|^s \right )
             + \ {\widetilde C_s \over \lambda^s}
            \sum_{<v,v'> \in \Gamma} |T_{v',v}|^s
            \E \left ( |G_{\Omega} (v, y;z) |^s
            \right ) \; ,
 \label{eq:fullbound}
 \end{multline}
with $\Gamma = \Gamma(W)$ the `cut-set' of $W$.
 \label{lem:fullbound}
\end{lem}

\begin{proof}
Starting from the first order
resolvent identity, \eq{eq:firstorder}, and taking expectation
values of its matrix elements, we find:
\begin{multline}
\E \left ( | G^{(\Gamma)}_{\Omega} (u,y;z)|^s \right ) \ \le \
    \E \left ( |G_{\Omega}(u,y;z)|^s \right ) \\
    + \sum_{<v,v'> \in \Gamma(W)} |T_{v',v}|^s
            \E \left ( | G^{(\Gamma)}_{\Omega}(u,v';z)|^s
            |G_{\Omega} (v, y;z) |^s
            \right ) \; ,
\end{multline}
where $\Gamma = \Gamma(W)$, and $G^{(\Gamma)}= G_{\Omega \backslash W}$.
It suffices, therefore, to show that in the last term the factor
 $| G^{(\Gamma)}_{\Omega}(u,v';z)|^s$ may be replaced (for an upper
 bound) by the constant ${\widetilde C_s \over \lambda^s}$.
This follows through a decoupling argument which we present in
the Appendix --- see
Lemma~\ref{lem:decouplinginequalities}.
\end{proof}

\begin{remark}
In the applications we shall use Lemmas~\ref{lem:diagram} and
\ref{lem:fullbound} both in the stated form and in the conjugated
form, with the arguments of the Green functions reversed.
One form of course implies the other (at conjugate energy).
\end{remark}

\masubsect{Proofs of the main results}

We are now ready to derive the results stated in the Introduction.
For simplicity these were stated in the context of the
Schr\"odinger operators, for which $T$ is the discrete Laplacian.
The proofs given in this section will be restricted to this case.
A more generally applicable treatment is presented in the next section.

\begin{proof_of}{Theorem \ref{thm:1}}
Assume that for some $z \in \C$ and a finite region $\Lambda$ the
smallness condition (\ref{eq:cond1}) holds. By
Lemma~\ref{lem:diagram} and translation invariance, we learn that
for any region $\Omega$ and any $x,y \in \Omega$ with $y \in \Z^d
\backslash \Lambda^{+}(x)$:
\begin{equation}
\E \left ( |G_\Omega(x,y;z)|^s
        \right )  \ \leq \ b \cdot {1 \over |\Gamma(\Lambda^+)|}
         \sum_{<v,v'> \in \Gamma(\Lambda^+(x))}
            \E \left (
                |G_{\Omega \backslash \Lambda^+(x)}(v',y;z)|^s
            \right ) \; ,
\label{eq:thm1iterate}
\end{equation}
where $b=b(\Lambda,z)$ of \eq{eq:cond1}, and $\Lambda(x)$ is the
translate of $\Lambda$ by $x$.

By Lemma~\ref{lem:1}, each of the terms in the sum is bounded by
$C_s/\lambda^s$.  Since the sum is normalized by the prefactor $
1/ |\Gamma(\Lambda^+)|$, the inequality (\ref{eq:thm1iterate})
permits to improve that bound for $\E (|G_\Omega(x,y;z)|^s)$  by
the factor $b \ (<1)$. Furthermore, the inequality may be iterated
a number of times, each iteration resulting in an additional
factor of $b$.

One should take note of the fact that the iterations bring in
Green functions corresponding to modified domains. It is for this
reason that the initial input assumption was required to hold for
modified geometries, {\it i.e.} not just for $\Lambda$ but also
for all its subsets.

Inequality~(\ref{eq:thm1iterate}) can be iterated
as long as the resulting sequences
($x, v', \ldots, v^{(n)}$) do not get closer to $y$
than the distance
$L= \sup\{ |u| \ | \ u \in \Lambda^+ \}$.
Thus:
\begin{equation}
\E \left ( |G_\Omega(x,y;z) |^s \right ) \ \le \ {C_s \over
\lambda^s} \cdot b^{\lfloor |x-y|/L \rfloor} \ \le \
{C_s \over \lambda^s b } \ e^{- \mu |x-y|}
\; , \label{eq:thm1optimalbound}
\end{equation}
with $\mu = |\ln b| / L$.
\end{proof_of}

Next, let us turn to the proof of the second theorem
(Thm~\ref{thm:2}). The main change is that we now proceed under
the assumption that the smallness condition holds for some region
$\Lambda$ without requiring it to hold also in all subsets.  As
explained in the introduction, the difference may be meaningful if
$H_{\omega}$ has extended boundary states in some geometry.

\begin{proof_of}{Theorem \ref{thm:2}}   Our first goal is to show
that under the assumption (\ref{eq:cond2}) there is $b< 1$ such
that for all pairs $\{x,y\}$ with $\Lambda(x) \subset \Omega$ and
$y \in \Omega \backslash \Lambda(x)$,
\begin{equation}
\label{eq:thm2iterate} \E
\left ( |G_\Omega(x,y;z)|^s \right ) \ \le \ b \
    \sum_{u \in \Lambda^+(x)  } P^l_x(u) \
         \E \left ( |G_\Omega(u,y;z)|^s \right ) \; ,
\end{equation}
with non-negative weights satisfying:
\begin{equation}
\sum_{u \in \Lambda^+(x)  } P^l_x(u) \ = \ 1 \; .
\label{eq:norm}
\end{equation}
We shall use this inequality along with its conjugate:
\begin{equation}
 \label{eq:reversedthm2iterate} \E
\left ( |G_\Omega(x,y;z)|^s \right ) \ \le \ b
    \sum_{v \in \Lambda^+(y)} P^r_y(v) \
         \E \left ( |G_\Omega(x,v;z)|^s \right ) \; ,
\end{equation}
where $ P^r_y(v)$ satisfy the suitable analog of the
normalization condition (\ref{eq:norm}).

It is important that -- unlike in the inequality
(\ref{eq:thm1iterate}), the functions which appear
on the right hand side of (\ref{eq:thm2iterate}) and
(\ref{eq:reversedthm2iterate}) are computed in the same
domain as those on the left hand side.

The first step is by Lemma~\ref{lem:diagram}, which yields
\begin{equation}
\E \left ( |G_\Omega(x,y;z)|^s \right ) \ \leq \
         \sum_ {<u,u'> \in \Gamma(\Lambda(x))}
         \gamma_x(<u,u'>)  \ \E \left (
         | G_{\Omega \backslash \Lambda(x)}(u',y;z)|^s
         \right ) \; ,
\end{equation}
whenever $\Lambda(x) \subset \Omega$ and $y \in \Z^d \backslash \Lambda(x)$,
with $\gamma_x(<u,u'>) $ specified in \eq{eq:gamma}.

Next, we apply Lemma~\ref{lem:fullbound}, \eq{eq:fullbound}, to bound
$\E \left (| G_{\Omega \backslash \Lambda(x)}(u',y;z)|^s
\right ) $ in terms of a sum of quantities of the form
$\E \left ( |G_{\Omega}(v,y;z)|^s \right )$
with  $v\in \Lambda^{+}(x)$.   The result is
initially expressed as a sum over bonds:
\begin{multline}
\E \left ( |G_\Omega(x,y;z)|^s \right ) \ \leq
    \sum_{<u,u'> \in \Gamma(\Lambda(x))}\gamma_x(<u,u'>) \
          \E \left (|G_\Omega(u',y;z)|^s \right )
          \\
    + \ {\widetilde C_s \over \lambda^s} \ \Theta
    \sum_{<u,u'> \in \Gamma(\Lambda(x))}
         \E \left (|G_\Omega(u,y;z)|^s \right ) \; ,
\label{eq:thm2fullbound}
\end{multline}
where, using translation invariance,
\begin{equation}
\nonumber \Theta \ := \ \sum_{<u,u'> \in \Gamma(\Lambda)}
\gamma_0(<u,u'>) \; .
\end{equation}

Collecting terms, and pulling out normalizing factors,
one may cast the inequality (\ref{eq:thm2fullbound}) in
the form (\ref{eq:thm2iterate}) with
\begin{eqnarray}
b  & := & \sum_{<u,u'> \in \Gamma(\Lambda(x))}
        \left( \gamma_x(<u,u'>)  \
        +  {\widetilde C_s \over \lambda^s} \Theta
         \right ) \ = \
       \left ( 1 + {\widetilde C_s \over \lambda^s}
        \ |\Gamma(\Lambda)| \right ) \ \Theta \\
    & = &  \left ( 1 + {\widetilde C_s \over \lambda^s}
        \ |\Gamma(\Lambda)| \right )^2
   \sum_{<u,u'> \in \Gamma(\Lambda)}
            \E \left ( | G_{\Lambda}(0,u;z)|^s
            \right )  \; .
\end{eqnarray}
The smallness condition (\ref{eq:cond2}) is nothing other than  the
assumption that $b  < 1$.

The above argument proves \eq{eq:thm2iterate}. By the
transposition, or time-reflection, symmetry of $H$ ($H^{T}=H$)
also \eq{eq:reversedthm2iterate} holds.
(Such symmetry of $H$ is not essential for our
analysis: it suffices to assume that the smallness condition
\eq{eq:cond2}  holds along with its transpose.)

We proceed in the proof by iterating the inequalities
(\ref{eq:thm2iterate}) and (\ref{eq:reversedthm2iterate}).  However
an adaptation is needed in the argument which was used in the proof of
Theorem~\ref{thm:1} since the iteration can be carried out only
as long as the two points (the arguments of the resolvent) stay at
distance $L= \sup \{ |u| : u \in \Lambda^{+}\}$ not only from
each other but also from the boundary  $\partial \Omega$.
The relevant observation is that for every pair of sites
$x,y\in \Omega$ there is a pair of integers $\{ n, m\} $ such that:
\begin{enumerate}
\item $n+m = \dist_{\Omega}(x,y)$ ,
\item
the ball of radius $n$ centered at $x$ and the ball of radius
$m$ centered at $y$ form a pair of disjoint subsets of $\Omega$.
\end{enumerate}
For the desired bound on $\E \left ( |G_\Omega(x,y;z) |^s \right
)$, we shall iterate  \eq{eq:thm2iterate}
$\lfloor n/L \rfloor$ times from the left,
and (\ref{eq:reversedthm2iterate}) $\lfloor m/L \rfloor$ times
from the right. Similar to
\eq{eq:thm1optimalbound}, we obtain:
\begin{equation}
\E \left ( |G_\Omega(x,y;z) |^s \right ) \ \le  \
{C_s \over \lambda^s b^2 } \ e^{- \mu \ \dist_{\Omega}(x,y)}
\; , \label{eq:thm2optimalbound}
\end{equation}
with $\mu = |\ln b| / L$.
\end{proof_of}

The third Theorem stated in the introduction  (Thm~\ref{thm:3}) is
the claim that the condition which is shown above to be sufficient
for exponential localization, in the sense of \eq{eq:fm}, is also a
necessary one. We shall now prove this to be the case.

\begin{proof_of}{Theorem~\ref{thm:3}}
Suppose that \eq{eq:fm} holds with some $A < \infty$ and $\mu >
0$. We need to show that also in finite systems the Green function
is sufficiently small between an interior point and the boundary.
To bound the finite volume function in terms of the infinite
volume one, we may use lemma~\ref{lem:fullbound}, by which
\begin{multline}
\sum_{<u,u'> \in \Gamma(\Lambda)} \E \left (|G_\Lambda (0,u;z)|^s
    \right ) \ \le \ \sum_{<u,u'> \in \Gamma(\Lambda)}\E \left (
    |G(0,u;z)|^s \right ) \\
+ \ {\widetilde C_s \over \lambda^s} \ |\Gamma(\Lambda)|
            \sum_{<v,v'> \in \Gamma(\Lambda)} |T_{v,v'}|^s \
            \E \left ( |G(0, v'; z) |^s
            \right ) \; ,
\label{eq:noname}
\end{multline}
for any finite region $\Lambda$ containing the origin.
We need to show that for  $\Lambda = [-L,L]^d$ with  $L$ large enough
\begin{equation}
 \left ( 1 + {\widetilde C_s \over
\lambda^s} |\Gamma(\Lambda)| \right )^2
    \sum_{<u,u'> \in \Gamma(\Lambda)}
            \E \left ( |G_\Lambda(0,u;z)|^s
            \right ) \ < \ 1 \; .
\label{eq:local}
\end{equation}
After applying \eq{eq:noname} to the terms on the left side of
 \eq{eq:local} we find that
the number of summands involved and their prefactors grow only
polynomially in $L$, whereas under our assumption the relevant
factors $ \E \left ( |G(0,u;z)|^s \right )$ are  exponentially
small in $L$.  Hence the condition (\ref{eq:local}) is satisfied
for $L$ large enough.
\end{proof_of}

\newpage

\masect{Generalizations} \label{sect:gen}

\masubsect{Formulation of the general results}

We shall now turn to some generalizations of the theorems which
were presented in Section~\ref{sect:main} for the random
Schr\"odinger operator. The setup may be extended in a number of
ways.

\begin{enumerate}
\item[1.] {\em Addition of magnetic fields.}
The hopping terms $\{T_{x,y}\}$ need not be real.
In particular, the present analysis remains valid when one
includes
in $H_{\omega}$ a constant magnetic field, or a random one with a
translation invariant distribution.
\end{enumerate}

A magnetic field  is incorporated in $T_{x,y}$ through  a factor
$\exp( -i  A_{x,y})$, with $A_{x,y}$ an anti-symmetric function of
the bonds.  (It represents the integral of the `vector potential'
$\times (-e/\hbar) $ along the bond $<x,y>$.) Except for the
trivial case, with such a factor $T$ is no longer
 shift invariant.
However, in the case of a constant magnetic field, $T$  will still
be invariant under appropriate``magnetic shifts'', which consist
of ordinary shifts followed by gauge transformations.

Translation-invariance plays a role in our discussion.  However,
since  gauge transformations do not affect the absolute values of
the resolvent, it suffices for us to assume that  $H_{\omega}$ is
{\em stochastically invariant under magnetic shifts} -- in the
sense of Definition~\ref{def:1}.

\begin{enumerate}
\item[2.] {\em Extended hopping terms.}
The discrete Laplacian may be replaced by an operator
with hopping terms of unlimited range.  For exponential
localization we shall however require $\{T_{x,y}\}$ to decay
exponentially in $|x-y|$.

\item[3.] {\em Off-diagonal disorder.}   $\{T_{x,y}\}$ may also
be made random.  It is convenient however to assume exponentially
decaying uniform bounds.  The regularity conditions on the
potential will now be assumed for the conditional distribution of
$V(x)$ at specified off-diagonal disorder.

\item[4.] {\em Periodicity.}  $H_{\omega}$ may also include a
periodic potential, i.e., \eq{eq:proto} may be modified to:
\begin{equation}
H_{\omega} \ = \  T_{x,y; \omega} + U_{per}(x) +
 \lambda V_{\omega}(x)
\end{equation}
This may be further generalized by requiring periodicity only of
the probability distribution of $H$.

\item[5.] {\em More general lattices.}
\end{enumerate}

In the previous discussion, the underlying sets $\Z^d$ may be
replaced by other graphs, with suitable symmetry groups.
The graph structure is relevant if the hopping terms
are limited to graph edges.
However, since we consider also operators
with hoping terms of unlimited range, let us formulate the
result for operators on $\ell^2(\T)$ where the underlying set
is of the form $\T = \G \times S $, with $\G$  a countable
group and $S$ a finite set.
We let $\dist(x,y)$ denote a metric on $\T$ which is
invariant under the natural action of  $\G$ on that set.

For example, this setup allows for $\T$ to be a Bethe lattice, or
a more general Cayley lattice. (Instructive discussion of some
statistical mechanical models in such settings may be found in
refs.\cite{BLPS}).  The set $S$ is included here in order to
leave room for periodic structures. We denote by   $\Cell$ the
``periodicity cell'', which is $\{\imath \}\times S$ where $\imath
$ is the identity in $\G$.

Some of the relevant concepts are summarized in the following
definition.

\begin{df}
\label{def:1} With  $\T = \G \times S$ as above, let  $H_{\omega}$
be a random operator on  $\ell^2(\T)$ (i.e., one with some
specified probability distribution), whose off-diagonal part is
denoted by $T_{\omega}$ and the diagonal part is referred to as
the potential (for consistency, we denote it as $\lambda
V_{\omega}$).
\begin{enumerate}
\item
We say that $H_{\omega}$ is
\underline{stochastically invariant under magnetic shifts}
if for each
$\kappa \in \G$ and almost every $\omega$ there is
a unitary map of the form
\begin{equation}
\left( U_{\kappa, \omega} \psi \right)(x) \ = \ e^{i \phi_{\kappa,
\omega}(x)  } \psi(\kappa x) \; ,
\end{equation}
(with some function $\phi_{\kappa, \omega}(\cdot)$ )
under which
\begin{equation}
U^{*}_{\kappa, \omega} \ H_{\omega} \ U_{\kappa, \omega} \
\stackrel{\mathcal D}{=}  \ H_{\omega} \; ,
\end{equation}
where $\stackrel{\mathcal D}{=} $ means equality of the probability
distributions.
\item
The operator is said to have
\underline{tempered off-diagonal matrix elements}, at a specified
value of $s<1$,
if  there is a kernel $\tau_{x,y}$, and some $m>0$, such that
$T_{x,y;\omega}\ \le \ \tau_{x,y}$, almost surely, and
\begin{equation}
\sup_{x\in \T} \
\sum_{y\in \T} \tau_{x,y}^s \ e^{+\, m \, \dist(x,y)} \ < \
\infty \; .
\label{eq:deftempered}
\end{equation}
\item We say that the potential  has
an  \underline{$s$-regular distribution} if for some $\tau > s$
the conditional distributions of $\{V_{\omega}(x)\}$, at specified
values of the hopping terms variables $\{T_{u,v;\omega}\}$, are
independent and satisfy the regularity conditions $R_1(\tau)$ and
$R_2(s)$ with uniform constants.
\end{enumerate}
\end{df}

Following is the generalization of Theorem~\ref{thm:1}.

\begin{thm}
Let $H_{\omega}$ be a random operator on $\ell^2(\T)$
($\T=\G\times S$, as above) with an $s$-regular distribution for
the potential $V_{\omega}(\cdot )$, and with tempered off-diagonal
matrix elements ($T_{x,y;\omega}$), which is stochastically
invariant under magnetic shifts. Assume that for some $z\in \C$
and a finite region $\Lambda \subset \T $, which contains the
periodicity cell $\Cell$,  the following is satisfied for all
subsets $W\subset \Lambda$
\begin{equation}
    \left (
        1 + {\widetilde C_s \over \lambda^s} \ \Xi_s(\Lambda)
    \right )
     \sup_{x\in \Cell}
     \sum_{<u,u'> \in \Lambda \times (\T  \backslash \Lambda)}
           \tau(u-u')^s \ \E\left (
                |<x| {1 \over
                    H_{W;\omega} - z
                    } |u>|^s
            \right ) \ < \  1 \; ,
\label{eq:thmgeneralcond2}
\end{equation}
where
\begin{equation}
\tau(v) = \sup_{u\in \T} \mbox{\rm ess sup}_{\omega}
     |T_{u,u+v;\omega}|
\; , \qquad
\Xi_s(\Lambda) \ = \  \sum_{<u,u'> \in \Lambda \times (\T
\backslash \Lambda)} \tau(u-u')^s \; .
\label{eq:tau}
\end{equation}
Then
there exist $\mu > 0$, $A< \infty$, such that for all
$\Omega \subset \T$, and all
$y\in \Omega$,
\begin{equation}
\sum_{x \in \Omega} \E_{\pm i 0} \left( |<x| {1 \over H_{\Omega;
\omega}-z } |y> |^s \right) \ e^{+  \, \mu \, \dist(x,y)} \ \le \
A \label{eq:summabledecay2}
\end{equation}
\label{thm:generalized1}
\end{thm}

\noindent{\bf Remarks:}

\noindent {\bf 1.} For graphs which grow at an exponential rate,
such as the Bethe lattice, exponentially decaying functions need
not be summable. The conclusion, \eq{eq:summabledecay2}, was
therefore formulated in the stronger form, which implies both
exponential decay, and almost sure summability. In particular, it
is useful to recall that for $s/2 < 1$:
\begin{equation}
\E\left( \left[ \sum_{y} |G(x,y)|^2 \right ]^{s/2} \right)  \ \le
\ \E\left(  \sum_{y} |G(x,y)|^{s} \right) \; .
\end{equation}

\noindent {\bf 2.} One may note  that in the more general theorem
we do make use of the ``decoupling Lemma'', which was not used in
Theorem~\ref{thm:1}.

\noindent {\bf 3.} Translation invariance played a limited role here:
the analysis extends readily to random operators with
non-translation invariant distributions, provided only that the
required bounds are satisfied uniformly for all translates of
$\Lambda$, and the distribution of the potential is
uniformly $s$-regular.  To demonstrate the required change
we cast the next statement in that form.

As we discussed in the preceding sections, condition
(\ref{eq:thmgeneralcond2}) may fail due to the existence of
extended states at some surfaces.  The following generalization of
Theorem~\ref{thm:2} provides criteria for localization in the bulk
which are less  affected by such surface states.

\begin{thm} Let $H_{\omega}$ be a random operator
 on $\ell^2(\T)$
($\T=\G\times S$, as above) with an $s$-regular distribution for
the potential $V_{\omega}(\cdot )$, and with tempered off-diagonal
matrix elements ($\{ T_{x,y;\omega} \} $).
Assume that for some $z\in \C$ and a
finite region $\Lambda$, $\Cell \subset \Lambda \subset \T$,
 \begin{equation}
    \left (
        1 + {\widetilde C_s \over \lambda^s} \ \Xi_s(\Lambda)
    \right )^2
    \sup_{x\in \T}
     \sum_{\substack {u \in \Lambda(x)  \\
                    u' \in  \T  \backslash \Lambda(x)} }
           \tau_{u,u'}^s \ \E \left (
                |<x| {1 \over
                    H_{\Lambda;\omega} - z [\bar z] } |u>|^s
            \right ) \ < \  1 \; ,
\label{eq:thmgeneralcond}
\end{equation}
where $\Lambda(x)$ is the unique translate of $\Lambda$, by an
element of $\G$, which contains $x$, and
$z [\bar z]$ means that the bound is satisfied for
both $z$ and $\bar z$.
Then the condition  (\ref{eq:summabledecay2})
holds for the full operator $H_{\omega}$ (i.e., with  $\Omega = \T$),
 and there  exist $B < \infty, \ \tilde \mu > 0$
with which for arbitrary  $\Omega \subset \T$:
 \begin{equation}
\E_{\pm i 0} \left( |<x| {1 \over  H_{\Omega;\omega}-z } |y> |^s
\right) \ \le \ B \  e^{- \, \tilde \mu \,  \dist_{\Omega}(x,y)}
\; . \label{eq:Omegadecay}
\end{equation}
\label{thm:generalized2}
\end{thm}

The modified distance $\dist_{\Omega}(x,y)$
is defined by the natural extension of  \eq{eq:dist}.

\masubsect{Derivation of the general results}

The derivation of Theorems \ref{thm:generalized1} and
\ref{thm:generalized2} follows very closely the proofs of
Section~\ref{sect:proof}.  The main difference is in the second
portion of the argument where we encounter a more general
``sub-harmonicity'' relation.

The
first part of the proof rests on  Lemmas \ref{lem:diagram} and
\ref{lem:fullbound} which are easily seen to extend to the setup
described in Theorem~\ref{thm:generalized2}.
(The hopping terms $T_{x,y}$ appearing in
section~\ref{sect:lemmas} are replaced with the uniform
upper-bound $\tau_{x,y}$.) We thus obtain the following
extension of the resolvent bounds.

\begin{lem} Let $H_{\omega}$ be a random operator
with the properties listed in Theorem~\ref{thm:generalized2}, and
let $\Lambda $ be a finite subset of $\T$, containing the periodicity cell
$\Cell$, for which the condition
(\ref{eq:thmgeneralcond2}) is   satisfied.
Then the following bound is valid for
any $x\in \Lambda, y\in \T \backslash \Lambda$,
\begin{equation}
\sup_{\Omega \subset \T} \E \left ( |G_\Omega(x,y;z)|^s \right ) \ \le \
 b \ \sum_{u \in \T}  p_\Lambda(x,u) \ \sup_{\Omega
\subset \T} \E \left ( |G_{\Omega} (u,y; z) |^s \right ) \; ,
\label{eq:lemma3.3a}
\end{equation}
with some $b < 1$ and a ``sub-probability kernel''
$p_\Lambda(x,u)$,  satisfying
\begin{equation} \sum_{u} p_\Lambda(x,u) \ \le \ 1 \; , \text{ and }
\sum_x p_\Lambda(x,u) \ \le \ 1 \; ,
\label{eq:sub-probability}
\end{equation}
which is tempered in the sense that for some $m>0$
\begin{equation}
\sup_x \sum_{u} e^{m \, \dist(x,u)} p_\Lambda(x,u) \ < \ \infty \;
, \text{ and } \sup_u \sum_x e^{m \, \dist(x,u)} p_\Lambda(x,u) \
< \ \infty \;.
\label{eq:tempered}
\end{equation}

Furthermore, assuming (\ref{eq:thmgeneralcond}) instead of
(\ref{eq:thmgeneralcond2}),  the following bound is valid for
any $x\in \Lambda, y\in \T \backslash \Lambda$, and $\Omega
\supset \Lambda  $
\begin{equation}
\E \left ( |G_\Omega(x,y;z)|^s \right ) \ \le \ \tilde b \
\sum_{u \in
\T} \widetilde p_\Lambda(x,u) \  \E \left ( |G_{\Omega } (u,y; z) |^s \right
) \; ,
\label{eq:lemma3.3b}
\end{equation}
with some $ \widetilde b < 1$ and $\widetilde p_\Lambda(x,u)$
which satisfies the same conditions as $p_\Lambda(x,u)$.
\label{lem:generalized2}
\end{lem}

The bounds presented in the above lemma may be read as  stating
that the resolvent $\E(|G(x,y;z)|^s)$ is sub-harmonic (we
use this term here in the sense of ``sub-mean'')  with respect
to a tempered probability kernel whenever $x,y$ are
sufficiently far apart. Theorems~\ref{thm:generalized2} and
\ref{thm:generalized1} follow from these bounds via a general
principle which applies to such sub-harmonic functions.  We expect
this principle to be well known, but for completeness we include a
proof here.

\begin{prop} Let $(\T, \dist)$ be a countable metric space,
$\Lambda \subset \T$  a finite subset,
and $g: \T \to \R $ a bounded and non-negative function, which
for all $x \in \T \backslash \Lambda$ satisfies:
 \begin{equation}
    g(x) \ \le \ b \sum_{u} p(x,u) g(u)  \;  ,
\label{eq:g}
\end{equation}
with  a kernel on $\T \times \T$ satisfying
\begin{equation}
\sup_x \sum_u   p(x,u) \ \le \ 1  \; ,
\qquad
\sup_u \sum_x   p(x,u) \ \le \  1 \; ,
\label{eq:kernel-prob}
\end{equation}
which is tempered in the sense of \eq{eq:tempered}.
Then $g(x)$ is exponentially  summable, i.e., for some $\mu > 0$:
\begin{equation}
\sum_y e^{\mu \, \dist(y,\Lambda) } \ g(y) \ < \ \infty \; .
\label{eq:summabledecay}
\end{equation}
\label{prop:subharmonic}
\end{prop}

\begin{proof}  One may read the claim as saying that
the function $g(\cdot)$ lies in the space $\ell^{1;\, \mu}(\T)$
of functions for which the following norm is finite:
\begin{equation}
\|f\|_{1, \mu} \ := \ \sum_{x \in \T} e^{\mu \, \dist (x, \Lambda)}
|f(x)| \; .
\end{equation}
We shall deduce this claim after arriving first at a bound
formulated within the larger  space of bounded functions
$\ell^{\infty}(\T)$.

Let  $P$ be the linear operator with the kernel $p(x,y)$.
Within $\ell^{\infty}(\T)$ the
operator acts as a contraction, since its norm there is
\begin{equation}
\| P \|_{\infty, \infty} \ = \ \sup_{x} \sum_{u} p(x,u) \ \le \ 1
\end{equation}
(using (\ref{eq:kernel-prob}) ).  It is convenient to paraphrase the
assumption on $g(\cdot )$ in the following form, which holds
for all $x\in \T$:
\begin{equation}
g(x) \ \le \ \|g\|_\infty \cdot I_\Lambda(x) \ + \ b \ [P \cdot
g](x) \; ,
\label{eq:globalbound}
\end{equation}
with $I_\Lambda$ the ``indicator function'' of $\Lambda$.
Iterating this relation $N$ times, one obtains a bound in the form of a
finite geometric series with a ``remainder'' which is uniformly
bounded by $ (b \  \| P\|_{\infty, \infty})^N \cdot  \|g\|_\infty $.
As  $N \to \infty$ the reminder vanishes, since
$(b \  \| P\|_{\infty, \infty}) < 1$, and one is left with a bound in
the form of a  convergent series:
\begin{equation}
g(x) \ \le \ \| g \|_\infty \ \sum_{n=0}^{\infty} b^n \
 [P^n \cdot I_\Lambda](x)   \; .
   \label{eq:seriesbound}
\end{equation}

We now note that for a finite region $\Lambda$, the function
$I_\Lambda$ lies in the ``weighted-$\ell^1$ space'' $\ell^{1;\, \mu}$.
The norm of  $P$ as an operator within $\ell^{1;\, \mu}$ is easily seen
to obey:
\begin{equation}
\| P \|_{1,\mu; 1, \mu} \ \le  \ \sup_u \sum_x e^{\mu \, \dist (x,u)}
p(x,u) \; .
\end{equation}
The expression on the right hand side is convex in $\mu$, and
by the temperedness assumption (the analog of \eq{eq:tempered})  it
is finite for small  enough $\mu > 0$.  Since convexity implies
continuity, using (\ref{eq:kernel-prob}) we conclude that there is
some $\mu > 0 $ for which
\begin{equation}
b \ \| P \|_{1,\mu; 1, \mu}\ < \ 1 \; .
\end{equation}
With this choice of $\mu$ we conclude:
\begin{equation}
\sum_{x} e^{\mu \, \dist(x, \Lambda)} \ g(x)  \ \equiv \
\| g \|_{1,\mu} \ \le \
{\| g \|_{\infty} \ | \Lambda | \over 1 - b \, \| P \|_{1,\mu; 1, \mu} }
 \ <  \ \infty  \; .
\end{equation}
\end{proof}

Theorems~\ref{thm:generalized1} and
\ref{thm:generalized2}  now follow by a combination  of the
proposition just shown with Lemma~\ref{lem:generalized2}.

\begin{proof_of}{Theorem \ref{thm:generalized1} }
To establish the claimed bound (\ref{eq:summabledecay2}) fix $y
\in \T$, and let $g(x) = \sup_\Omega \E(|G_\Omega(x,y;z)|^s)$. We
note that for each $x\in \T$ there is a unique element of the
symmetry group, $h_x \in \G$, such that $h_x x \in \Lambda$.
Starting from the kernel $p_\Lambda(h_x x,h_x u)$ which appears in
Lemma~\ref{lem:generalized2}, let us define a shift-invariant
kernel $p(x,y)$ by:
\begin{equation}
p(x,u) \ = \ p_\Lambda(h_x x,h_x u) \; .
\label{eq:pxy}
\end{equation}
Due to the shift invariance of the distribution of $H_{\omega}$,
\eq{eq:lemma3.3a} implies that the function $g(x)$
 is sub-harmonic, in the sense of (\ref{eq:g}), with respect to
the kernel $p(x,u)$, which satisfies (\ref{eq:kernel-prob}) and is
tempered . Thus, a direct application of
Proposition~\ref{prop:subharmonic} yields now the claimed bound
(\ref{eq:summabledecay2}).
\end{proof_of}

\begin{proof_of}{Theorem \ref{thm:generalized2} }
The situation to be discussed now is different from that
encountered in the last proof in
that now for each $\Omega$ the basic sub-harmonicity bound
can be assumed only for points which are not too
close to the boundary $\partial \Omega$.
The claim made for the special case $\Omega = \T $ is covered by the
above analysis.
However, the  second claim, i.e., \eq{eq:Omegadecay}, requires a somewhat
different argument.

The argument we shall use shadows the proof of
Proposition~\ref{prop:subharmonic}, replacing there the
weighted-$\ell^{1}$ estimate by its weighted-$\ell^{\infty}$
version. The starting observation is that $\E (
|G_\Omega(x,y;z)|^s )$ has the sub-mean property with respect to
averages over either $x$ or $y$ -- provided the point is at
distance at least  ${\rm diam}(\Omega)$ from the other and from
the boundary $\partial \Omega$. (In allowing the averaging
procedure to occur from either side, we rely on the fact that  the
smallness condition holds for both the kernel $G(x,y;z)$ and its
conjugate, or equivalently the fact that the smallness condition
is assumed to hold for both $z$ and $\bar z$ .)

To cast the situation in terms reminiscent of the proof of
Proposition~\ref{prop:subharmonic}, let us  consider the function
$g(<x,y>)= \E ( |G_\Omega(x,y;z)|^s )$ as defined over the space
of pairs, $\Omega \times \Omega$, equipped with the distance
function
\begin{equation}
\dist_\Omega(<x_1,y_1>,<x_2,y_2>) \ = \ \dist_\Omega(x_1,x_2) +
\dist_\Omega(y_1,y_2) \; .
\end{equation}
For $<x,y>$ not in the set
$W :=  \{ <u,v> \ | \ \dist_\Omega(u,v) \le 2 L$,
with $L= {\rm diam}(\Lambda) \} $,
we have the basic sub-mean estimate:
\begin{equation}
g(<x,y>) \ \le \ b \ \sum_{<u,v>} \widetilde p(<x,y>,<u,v>) \ g(<u,v>) \; ,
\end{equation}
with
\begin{equation}
\widetilde p(<x,y>,<u,v>) :=
\left\{
  \begin{array}{ll}
       p(x,u) \, \delta_{y,v} &
          \mbox{\footnotesize if
              \ $\dist_{\Omega}(x,y) >  2 L$ and $\dist(x,\partial
              \Omega) >  L$  } \; , \\
            \delta_{x,u} \, p(y,v)  &
                 \mbox{\footnotesize if
                \   $\dist_{\Omega}(x,y) > 2 L$ and $\dist(x,\partial
                  \Omega) \le L$  }  \; , \\
            \delta_{x,u} \, \delta_{y,v} &
                \mbox{\footnotesize if \  $\dist_{\Omega}(x,y) \le 2
                L$   }       \; ,
          \end{array}
          \right.
   \end{equation}
where $p(x,y)$ is given by \eq{eq:pxy}.

By repeating the arguments seen there we find that
$g(<x,y>)$ obeys the analog of \eq{eq:seriesbound} ---
formulated within the space $\ell^{\infty}( \Omega \times \Omega)$,
 with the set $\Lambda$ replaced by $W$, and the operator $P$ replaced
by $\widetilde P$ defined by the kernel $\widetilde p(<x,y>,<u,v>)$.
Unlike in the previous case, we have no fixed bound on the size of
 the set $W$.  Thus we shall not use here   the weighted-$\ell^{1}$
estimate.  However, we may reuse the argument applying it to
weighted-$\ell^{\infty}$  norm   of $g(\cdot)$, which is defined
as:
\begin{equation}
\| g \|_{\infty; \mu} \ = \ \sup_{<x,y>}  e^{\mu \, \dist(x,y)} \, |g(<x,y>)|
\end{equation}
The conclusion is that there is some $\mu > 0$ at
which $\| g \|_{\infty; \mu} <
\infty$.  Equivalently:
\begin{equation}
\E \left ( |G_\Omega(x,y;z)|^s \right ) \ \le \ \c e^{-\mu \,
\dist_\Omega(x,y)} \; \; ,
\end{equation}
as claimed in Theorem~\ref{thm:generalized2}.
\end{proof_of}

\masect{Some Implications}
\label{sect:implications}

We shall now present a number of implications of the finite volume
criteria for localization, focusing on the finite dimensional
lattices $\Z^d$. The  statements will bear some resemblance to
results derived using the multiscale approach, however the
conclusions drawn here go beyond the latter by yielding results on
the exponential decay of the \underline{mean values}.  The
significance of that was described in the introduction.

\masubsect{Fast power decay $\Rightarrow$ exponential decay}

An interesting and useful implication
(as is seen below)
is that fast enough power law implies exponential decay.
In this sense,  random Schr\"odinger operators join
other statistical mechanical models in which
such principles have been previously recognized.
The list includes the general
Dobrushin-Shlosman results~\cite{DoSh}  and the more specific two-point
function bounds in: percolation
(Hammersley\cite{Ham} and Aizenman-Newman~\cite{AiNew}),
Ising ferromagnets
(Simon~\cite{Simon} and Lieb~\cite{Lieb}),
certain  $O(N)$ models (Aizenman-Simon~\cite{AiSi}),
and  time-evolution models (Aizenman-Holley~\cite{AiHo},
Maes-Shlosman~\cite{MaSh}.)

\begin{thm}  Let
$H_{\omega}$ be a random operator on $\ell^2(\Z^d)$ with an
$s$-regular distribution for the potential ($V_{\omega}(x)$) and
tempered off-diagonal matrix elements ($T_{x,y;\omega}$). There
are $L_0, B_1, B_2 <\infty$, which depend only on the temperedness
bound (\ref{eq:deftempered}), such that if  for some $E\in \R$ and
some finite  $L \ge L_0 $,
 either
\begin{equation}
           L^{3(d-1)} \ \sup_{\ L / 2 \le \|x-y\| \le L}
            \E \left(     |<x| {1 \over
                    H_{\Lambda_L(x), \omega} - E} |y>|^s
            \right ) \ \le \ B_1 \; ,
\label{eq:powerlaw1}
\end{equation}
or
\begin{equation}
     L^{4(d-1)} \ \sup_{\ L / 2 \le \|x-y\| \le L}
      \E \left(
                |<x| {1 \over
                    H_{\omega} - E - i 0} |y>|^s
            \right ) \ \le \ B_2 \; ,
\label{eq:powerlaw2}
\end{equation}
where $\Lambda_L(x)=[-L,L]^d+x$ and $\|y\| \equiv \max_j |y_j|$,
then the \underline{exponential} localization (\ref{eq:fm}) holds
for all energies in some open interval $(a,b)$ containing $E$.
\label{thm:power=>exp}
\end{thm}

\begin{proof}
By Theorem~\ref{thm:generalized2},
to establish exponential decay at the energy $E$ it suffices to show
 that for each $x\in \Z^d$
\begin{equation}
 \left (
        1 + {\widetilde C_s \over \lambda^s} \ \Xi_s(\Lambda_L)
    \right )^2
    \sum_{\substack {u  \in \Lambda_L(x) \\ u' \in \Z^d \backslash
\Lambda_L(x) } }
               \tau_{u,u'}^s \ \E \left (
                |G_{\Lambda_L(x)}(x,u;E)|^s
            \right ) \ < \  1 \; .
\label{eq:suff}
\end{equation}
Because the off diagonal elements are tempered
we have the following bounds
\begin{equation}
\tau_{u,u'}^s \ \le \  \c  \ e^{-m |u-u'|} \; , \qquad
\Xi_s(\Lambda_L) \ \le \  \c  \  L^{d-1} \; ,
\end{equation}
for some $m > 0$, and all $L > 1 $.
Under the assumption \eq{eq:powerlaw1}:
\begin{multline}
  \sum_{\substack {u  \in \Lambda_L(x) \\ u' \in \Z^d \backslash
\Lambda_L(x)}}
               \tau_{u,u'}^s \ \E \left (
                |G_{\Lambda_L(x)}(x,u;E)|^s
            \right ) \ \le \  \\
  \le  \   {\widetilde C_s \over \lambda^s}\  \c \ (L/2)^d \
  e^{-mL}/2  \  + \  \qquad  \  \\
     + \ \c \ \sup_{\ L / 2 \le \|x-y\| \le L}
      \E \left(  |<x| {1 \over   H_{\Lambda_L(x), \omega} - E} |y>|^s
\right )
      \  L^{d-1}  \; .
\label{eq:tiddledoo}
    \end{multline}
For this bound the sum was split according to
 $ \|u-u'\| < ({\rm or} \  \ge)  L/2$, and in the first case
 we used the uniform upper bound
$\E( |G(x,u;E)|^s) \le   \widetilde C_s /  \lambda^s$.

It is now easy to see that with an appropriate choice of $L_0$ and
$B_1$  condition (\ref{eq:powerlaw1}) implies the claimed bound
(\ref{eq:suff}) -- for the given energy $E$.  The extension to an
interval of energies around $E$ then follows from  the continuity
of the fractional moments of \underline{finite volume} Green
functions.

To show the sufficiency of the second condition, we first use
Lemma~\ref{lem:fullbound} to  bound finite
volume Green functions in terms of the corresponding
infinite volume funtions
\begin{equation}
 \E \left ( | G_{\Lambda_L(x)}(x,y;E)|^s \right )
  \le   \ \E  \left (| G( x, y;E)|^s \right )
    + \ {\widetilde C_s \over \lambda^s}  \sum_{\substack{
                        u \in \Lambda_L(x) \\ u' \in  \Z^d
                        \backslash \Lambda_L(x) }}
                    \tau_{u',u}^s \ \E  \left
                    ( | G( x, u';E)|^s \right ) \; .
\end{equation}
Splitting the sum as in \eq{eq:tiddledoo}, we get
\begin{multline}
\sup_{\ L / 2 \le \|x-y\| \le L}
      \E \left ( | G_{\Lambda_L(x)}(x,y;E)|^s \right ) \  \le  \\
     \le    \left[ {\widetilde C_s \over \lambda^s}\ \right]^2 \c \ (L/2)^d \
  e^{-mL}/2  \  + \ \qquad  \    \\
  + \  \left( 1 + \c \ L^{d-1} \right) \ \times \ L^{d-1} \
   \sup_{\ L / 2 \le \|x-y\| \le L}
   \E \left ( | G(x,y;E)|^s \right )
\label{eq:tiddledo2}
\end{multline}
The combination of \eq {eq:tiddledo2} with (\ref{eq:tiddledoo}),
yields the claim - for the given energy.  Again, the existence of an open
interval of energies in which the condition is met
is implied by the continuity of the finite-volume  expectation
values.
\end{proof}

\masubsect{Lower bounds for $G_{\omega}(x,y; E_{{\rm edge} }+i0)$
at mobility edges}

Boundary points of the continuous spectrum  are often referred to
as {\em mobility edges}.  (In an ergodic setting the location of
such points does not depend on the realization $\omega$
\cite{KuSu}.) The proof of the occurance  of continuous spectrum
for random stochastically  shift-invariant operators on $\Z^d$ is
still an open problem (one may add that we are glossing here over
some fine distinctions in the dynamical behaviour
\cite{dynamics}). However it is intersting to note that
Theorem~\ref{thm:power=>exp} directly yields the following pair of
lower bounds on the decay rate of the Green function at mobility
edges, $E_{{\rm edge} }$, for stochastically shift invariant
random operators with regular probability distribution of the
potential:
\begin{equation}
    \ \sup_{\ L / 2 \le \|y\| \le L}
      \E \left(
                |<0| {1 \over
                    H_{[-L,L]^d, \omega} - E_{{\rm edge} } } |y>|^s
            \right ) \ \ge \ B_1 \  L^{-3(d-1)} \; ,
\label{eq:mobilityedge1}
\end{equation}
\begin{equation}
    \ \sup_{\ L / 2 \le \|y\| \le L}
      \E \left(
                |<0| {1 \over
                    H_{\omega} - E_{{\rm edge} } - i 0} |y>|^s
            \right ) \ \ge \ B_2 \  L^{-4(d-1)} \; ,
\label{eq:mobilityedge2}
\end{equation}
with $\|y\| \equiv \max_{j} |y_j|$.
 We do not expect the power
laws provided here to be optimal. As mentioned above, vaguely
similar bounds are known for the critical two-point functions in
certain statistical mechanical models (percolation, Ising spin
systems, and some $O(N)$ spin models).

\masubsect{Extending off the real axis}

For various applications, such as the decay of the projection
kernel (see  \cite{AG} Sect. 5), it is useful to have bounds
on the resolvent at $z=E+i\eta$ which are uniform in $\eta$.
The following result shows that in order to establish such
uniform bounds it is sufficient to verify our criteria
for real energies in some neighborhood of $E$.

\begin{thm}  Let
$H_{\omega}$ be a random operator on $\ell^2(\Z^d)$
with an $s$-regular distribution for the potential
($V_{\omega}(x)$)
and  tempered off-diagonal matrix elements ($T_{x,y;\omega}$).
Suppose that for some $E\in \R$, and $\Delta E > 0$, the
following bound holds uniformly for
$\xi \in [E-\Delta E, E+\Delta E ]$:
\begin{equation}
    \E \left(  |<x| {1 \over H_{\omega} - \xi - i 0} |y>|^s
            \right ) \ \le \ A\,  e^{-\mu |x-y|}  \; .
\label{eq:real}
\end{equation}
Then for all $\eta \in \R$:
\begin{equation}
    \E \left(  |<x| {1 \over H_{\omega} - E - i \eta} |y>|^s
            \right ) \
 \le \ \widetilde A\ \ e^{-\tilde \mu  |x-y|}   \; ,
\label{eq:eq:strip}
\end{equation}
with some $\widetilde A  < \infty$ and $\tilde \mu > 0$ --
which depend on  $\Delta E$
and  the bound (\ref{eq:real}).
\label{thm:strip}
\end{thm}

\noindent{\bf Remarks:}

\noindent  {\bf 1.} This result is not needed in situations
covered by the \underline{single site} version of the criterion
provided by Theorem~\ref{thm:1}, since if \eq{eq:singlesite} is
satisfied at some $E\in \R$ then it automatically holds uniformly
along the entire line $E + i\R$. We do not see a monotonicity
argument for such a deduction in case of other finite-volumes.

\noindent {\bf 2.} One way to derive the statement is by using the
fact that exponential decay may be tested in finite volumes: if a
finite volume criterion holds for some $E$ then continuity allows
one to extend it to all $E + i \eta$ with $\eta$ sufficiently
small. The Combes-Thomas estimate~\cite{CT} can then be used to
cover the rest of the line $E+i\R$. However, by this approach one
gets only a weaker decay rate for energies off the real axis.   It
is tempting to think that some contour integration argument could
be found to significantly improve on that.  The proof given below
is a step in that direction (though it still leaves one with the
feeling that a more efficient argument should be possible).

\begin{proof}
Assume that the condition (\ref{eq:real}) is satisfied for all
$\xi \in [E-\Delta E, E+ \Delta E]$.
We shall show that this implies that for any power $\alpha$
\begin{equation}
    \E \left(  |<x| {1 \over H_{\omega} - \xi - i \eta} |y>|^s
            \right ) \ \le \
             { A_{ \alpha} \over |x-y|^{ \alpha} }\;  ,
\label{eq:power2}
\end{equation}
with the constant $A_{\alpha}< \infty$ uniform in $\eta$.
The stated conclusion then follows by
an application of  Theorem~\ref{thm:power=>exp} (and the
uniform  bounds seen in its proof).

We shall deal separately
with  large and small  $|\eta|$, splitting the two regimes at
$\Delta E \times \pi/ \alpha $.
 The case $|\eta| \ge \Delta E \times \pi/ \alpha $
is covered by the
general bound of Combes-Thomas~\cite{CT}, which states that:
\begin{equation}
|G(x,y;E+{\rm i}\eta)| \le (2/\eta){\rm e}^{-m|x-y|}
\label{eq:CT}
\end{equation}
for any $m\ge 0$ such that
\begin{equation}
\sum_{x\in \Z^d} \tau(x) \, (e^{m|x|}-1) \ \le \ \eta/2 \; .
\end{equation}

To estimate the resolvent for
 $|\eta| \le \Delta E \times \pi/ \alpha $, we shall use the
fact that  the  function
\begin{equation}
f_L(\zeta) \ = \   \E \left( |G_{[-L,L]^d}(x,y; \zeta )|^s \right)
\end{equation}
is subharmonic in the upper half plane, and continuous
at the boundary.  The subharmonicity is a general consequence of
the analyticity of the resolvent in $\zeta$, and the continuity
is implied through the continuity of the distribution
of the potential.  $L$ serves as a convenient cutoff, which may be
removed after the bounds are derived (since
$H_{[-L,L]^d, \omega} \ \too{L\to \infty} H_{\omega}$
in the strong resolvent sense).

Let $D\subset \C$ be the triangular region in the upper half plane
in the form of an equilateral triangle based on the real interval
$[E-\Delta E, E + \Delta E]$ with the side angles equal to $
\theta$ -- determined by the condition
\begin{equation}
\alpha \ = \ {2 \pi \over \theta } - 1 \; .
\label{eq:theta}
\end{equation}
The Poisson-kernel representation of harmonic
functions yields, for $E+ i \eta \in D$,
\begin{equation}
f_L(E + i \eta ) \ \le  \ \int_{\partial D} f_L(\zeta)
\ P^{D}_{E+ i \eta}(d \zeta)
\end{equation}
where $P^{D}_{E+ i \eta}(d \zeta) $ is a certain
probability measure on $\partial D$.
We now rely on the fact that this probability measure satisfies
\begin{equation}
P^{D}_{E+ i \eta}(d \zeta) \ \le \ \c \  d(\eta^{2 \pi / \theta}) \, /
\Delta E^{2 \pi / \theta} \; .
\end{equation}
(This is easily understood upon the
unfolding of $D$ by the map
$z \mapsto z ^{2 \pi / \theta}$
applied from either of the base corners of $D$,
i.e., from $\zeta = E \pm \Delta E$, and a  comparison with
the Poisson kernel in the upper half plane.)

For $\zeta \in \partial D \cap \R$ the integrand
satisfies the exponential  bound (\ref{eq:real}).
Along the rest of the boundary of $D$ we use the Combes-Thomas bound
(\ref{eq:CT}).
Putting it all together we get
\begin{equation}
f_L(E + i \eta ) \ \le \  A\,  e^{-\mu |x-y|} \ + \
\c  \int_{0}^{\Delta E\,  \theta }
{2 \over \eta } \, e^{- \c\, |x-y| \, \eta}
\ d(\eta^{2 \pi / \theta})\ /
\Delta E^{2 \pi / \theta}  \; .
\end{equation}
The claimed \eq{eq:power2} follows by simple integration, and the
relation (\ref{eq:theta}).
\end{proof}

\masubsect{Relation with the multiscale analysis and density of
states estimates}

Using the above results we shall now show that the fractional
moment localization condition is satisfied throughout the regime
for which localization can be shown via the multiscale analysis,
and also in regimes over which one has suitable bounds (e.g., via
Lifshitz tail estimates) on the density
of states of the operators restricted to finite regions
$\Lambda_{L} = [-L,L]^d$.  The following result is useful for
the latter case.

\begin{thm} Let $H_{\omega}$ be a random operator on $\ell^2(\Z^d)$
with tempered off-diagonal matrix elements ($T_{x,y;\omega}$) and
a distribution of the potential which is $s$-regular for all $s$
small enough, which is stochastically invariant under magnetic
shifts.  Then, given $\beta \in (0,1)$, $C_1 > 0$, and  $\xi > 3 (d-1)$,
there exist $L_0>0$ and $C_2 > 0$ such that
if for some $L \ge L_0$
\begin{equation}
\P \left [ \dist \left ( \sigma(H_{\Lambda_{L}; \omega}), E
\right) \le C_1  L^{-\beta} \right ] \ < \ C_2 L^{-\xi} \; ,
\label{eq:finitecondition}
\end{equation}
at some energy $E$, then the exponential
localization condition (\ref{eq:fm}) holds in some open interval
containing $E$. \label{thm:tails}
\end{thm}

The condition (\ref{eq:finitecondition}) is similar to the one
used in the multiscale analysis, although there one can also find
a sufficient diagnostic with arbitrary $\xi>0$. It may therefore
not be initially clear that the methods of this paper may be used
throughout the regime in which the multiscale analysis applies.
However, the proof of Theorem~\ref{thm:tails} is easily adapted to
prove the following result which implies fractional moment
localization via the {\em conclusions} of the multiscale analysis.

\begin{thm} Let $H_\omega$ be a random operator with tempered
off-diagonal matrix elements ($T_{x,y;\omega}$) and a distribution
of the potential which is $s$-regular for all $s$ small enough,
which is stochastically invariant under magnetic shifts.  If for
some $E \in \R$ there exist $A < \infty$, $\mu > 0$ , and $\xi >
3(d-1)$ such that
\begin{equation}
\lim_{L \rightarrow \infty} L^\xi \P \left [ |G_{\Lambda_L;\omega}
(0,x) | \ > \ A e^{-\mu |x|} \rm{~for~some~}x  \in \Lambda_L
\right ] \ = \ 0 \; , \label{eq:multiscale}
\end{equation}
then the exponential localization condition (\ref{eq:fm}) holds in
some open interval containing $E$. \label{thm:multiscale}
\end{thm}

\noindent{\bf Remarks:}

\noindent {\bf 1.} When the multiscale analysis applies, it allows one
to conclude that there are $A < \infty$ and $\mu > 0$ such
that the probabilities appearing on the left side of
\eq{eq:multiscale} decay faster than {\em any} power of $L$ as $L
\rightarrow \infty$. Thus, the conclusions of the multiscale
analysis imply that exponential localization in the stronger sense
discussed in our work applies throughout the regime which may be
reached by this prior method.

\noindent {\bf 2.}  It is of interest to combine the criterion
presented above with Lifshitz tail estimates on the density of
states at the bottom of the spectrum, $E_0$, and at band edges.
Using Lifshitz tail estimates, it is possible to show that
\cite{Si85}:
\begin{equation}
\P \left [ \inf \sigma(H_{\Lambda_{L}; \omega}) \le E_0 + \Delta E
\right ] \ \le \ \c \ L^d e^{-\Delta E^{-d/2}} \; .
\label{eq:bottom}
\end{equation}
Theorem~\ref{thm:tails} then implies fractional moment
localization in a neighborhood of $E_0$; we need only choose
$\Delta E \propto L^{-\beta}$ with $\beta \in (0,1)$ for large
enough $L$. Previous results in this vein may be found in
\cite{FiKl,BCH,KSS,Stollmann}.

\begin{proof_of}{Theorems~\ref{thm:tails} and
\ref{thm:multiscale}}
We first prove Theorem~\ref{thm:tails} and then indicate
how the proof can be modified to show Theorem~\ref{thm:multiscale}.

Fix an energy $E \in \R$.
For $L > 0$, define
\begin{equation}
p_L(\delta) \ := \ \P \left [ \dist \left ( \sigma(H_{\Lambda_{L};
\omega}), E \right) \le \delta \right ] \; ,
\end{equation}
and let
\begin{equation}
\delta_L \ := \ C_1 L^{-\beta} \;.
\end{equation}
We will show that for suitable $s\in(0,1)$, $L_0 >0$
and $C_2 > 0$, if
\begin{equation}
p_L(\delta_L) \ < \ C_2 L^{-\xi}
\end{equation}
then the input condition (\ref{eq:powerlaw1}) of
Theorem~\ref{thm:power=>exp}:
\begin{equation}
           L^{3(d-1)} \ \sup_{\ L / 2 \le \|y\| \le L}
            \E \left(     |<0| {1 \over
                    H_{\Lambda_L, \omega} - \widetilde E} |y>|^s
            \right ) \ \le \ B_1 \; ,
\label{eq:indiansummer}
\end{equation}
is satisfied for all energies $\widetilde E \in [E-\half \delta_L,
E + \half \delta_L]$. Exponential localization in the
corresponding interval (and strip, with $\eta \ne 0$) follows then
by Theorems~\ref{thm:power=>exp} (and Theorem~\ref{thm:strip}).

First we must show how to estimate $\E \left (
|G_{\Lambda_L;\omega}(0,u; \widetilde E)|^s \right )$ in terms of
$p_L(\delta)$.  This is achieved by considering separately the
contributions from the ``good set'':
\begin{equation}
\Omega_G \ = \ \{ \omega \ | \ \dist \left (
\sigma(H_{\Lambda_{L}; \omega}), E \right) > \delta \} \; ,
\end{equation}
and its complement, the ``bad set'':
$\Omega_B \ = \ \Omega_G^c$.

On the ``good set'',  $\omega \in \Omega_G$, the energy
$\widetilde E$ is at a small yet
significant distance ($\Delta E \ge \half \delta $) from the
spectrum of $H_{\Lambda_L;\omega}$.  In this situation, we use
the Combes-Thomas~\cite{CT} bound, by which:
\begin{equation}
|G_{\Lambda_L; \omega}(0,u; \widetilde E)|
    \ \le \ {2 \over  \Delta E} e^{-{1 \over 2} \Delta E
    |u|} \; .
\end{equation}
The above estimate does not apply on the ``bad set''. However,
using the H\"older inequality, we find that the net contribution
to the expectation is small because $\P(\Omega_B)=p_L(\delta)$ is
small.  The two estimates are combined in the following bound:
\begin{multline}
\E \left ( |G_{\Lambda_L; \omega}(0,u;\widetilde E)|^s \right ) \\
\begin{split}
 = & \ \E \left (   |G_{\Lambda_L; \omega}(0,u;\widetilde E)|^s
                    \ I[\omega \in \Omega_G] \right )
    \ + \ \E \left ( |G_{\Lambda_L; \omega}(0,u;\widetilde E)|^s
                    \ I[\omega \in \Omega_B] \right ) \\
 \le & \ {4^s \delta^{-s }} e^{- s \, |u|\,  \delta \, / 4 } \
    + \ \E \left (|G_{\Lambda_L; \omega}(0,u;\widetilde E)|^t \right  )^{s
\over t}
    \ \E \left ( I[\omega \in \Omega_B]  \right )^{1 -
    {s \over t} } \\
 \le & \ {4^s \delta^{-s}} e^{- s \, |u|\,  \delta \, / 4 } \
  + \  C_t^{s \over t}/ \lambda^s \, p_L(\delta)^{1 - {s \over
    t}} \; ,
 \end{split}
 \end{multline}
where  $t$ is any number greater than $s $ for which the
distribution of the potential is still $t$-regular (i.e., $C_t <
\infty$).

The required bound, \eq{eq:indiansummer}, is satisfied once one chooses
$s$ small enough so that
$\xi \ge {t \over t-s} 3 (d-1)$,
and $L_0$ large enough so that  for $L > L_0$
\begin{equation}
4^sC_1^{-s} L^{3(d-1)-s\beta}
     e^{-s \,C_1 \,  L^{1-\beta} \, / 4} \ \le \  B_1/2   \; .
\end{equation}

Finally let us remark on how this argument can be adapted to prove
Theorem~\ref{thm:multiscale}.  We simply define the good and bad
sets differently:
\begin{equation}
\Omega_G \ = \ \{ \omega \ | \ |G_{\Lambda_L;\omega}(0,x)| \ \le \
A e^{-\mu |x|} \rm{~for~all~} x \in \Lambda_L \} \; ,
\end{equation}
and $ \Omega_B \ = \ \Omega_G^c $ , and then proceed as in the
proof of Theorem~\ref{thm:tails} using H\"older's inequality to
estimate the contributions from $\Omega_B$.  It is easy to see
that for large $L$, the condition (\ref{eq:multiscale}) implies
that the input for Theorem~\ref{thm:power=>exp} is satisfied.
\end{proof_of}

Thus, we have seen here that the fractional moment localization
condition holds throughout the regime for which loclization can be
established by any available methods.  This is meaningful since
that condition carries a number of physically significant
implications.

\newpage
\startappendix

\maappendix{Dynamical Localization} \label{sect:dynamical}

Among the implications of the fractional moment condition is
dynamical localization, expressed through uniform exponential
decay of the average time evolution kernels:
\begin{equation}\label{eq:dynamicallocalization}
 \E \left ( \sup_{t \in \R} |<x| P_{H_\omega \in F} \ e^{itH}
               |y>| \right ) \ \le \ A \ e^{-\mu
 |x-y|} \;,
\end{equation}
where $P_{H_\omega \in F}$ indicates the spectral projection of
$H_\omega$ onto a set $F \subset \R$ in which the fractional
moment condition is known to hold.  A derivation of this
implication, under some auxiliary assumptions on the distribution
of the potential, was given in ref.~\cite{Ai94}. For
completeness we offer here a streamlined version of that argument,
which also extends the result in that we now allow $F$ to be an
unbounded set (in particular the full real line).

The inequality expressed in \eq{eq:dynamicallocalization} is not
special to the time evolution operators $f_t(E)=e^{itE}$; it follows, rather,
from a similar bound on the average total mass of the spectral
measures, $\mu^{x,y}_\omega$, associated to {\em pairs} of sites
$x,y$.  The measures are defined by the spectral representation:
\begin{equation}
\int f(E) \mu^{x,y}_\omega(dE) \ := \ <x|f(H_\omega)|y> \; ,
\end{equation}
for bounded Borel functions $f$.  In the following discussion
we denote by $|\mu^{x,y}_\omega|$ the {\em absolute
value} (sometimes called the {\em total variation}) of
$\mu^{x,y}_\omega$.

\begin{thm} Let $H_\omega$ be a
random operator on $\ell^2(\Z^d)$ with tempered off-diagonal
matrix elements and a potential $V_\omega$ which satisfies:
\begin{enumerate}
\item For some $\delta \in (0,1)$, the $\delta$-moments of $V_\omega$,
$\E \left ( |V_\omega(x)|^\delta \right )$, are uniformly bounded.
\item For each $x \in \Z^d$ the conditional distribution of
$v=V_\omega(x)$ at specified values of all other matrix elements
has a density $\rho^x_\omega(v)$, and the functions
$\rho^x_\omega$ are uniformly bounded.
\end{enumerate}
Suppose there is an energy domain $F \subset \R$ on which
$H_\omega$ satisfies a uniform fractional moment bound, {\it
i.e.}, there exist $A < \infty$ and $\mu > 0$ such that, for some
$s \in (0,1)$,
\begin{equation}\label{eq:fracmomcond}
\E \left ( | <x| {1 \over H_{\Lambda;\omega} -E} |y> |^s \right )
\ \le \ A \ e^{-\mu |x,y|} \; ,
\end{equation}
for any finite region $\Lambda \subset \Z^d$, any pair of sites
$x,y \in \Lambda$, and every $E \in F$. Then there exist $A' <
\infty$ and $\mu' > 0$ such that for any pair of sites $x,y \in
\Z^d$,
\begin{equation} \label{eq:spectraldecay}
\E \left ( |\mu^{x,y}_\omega|(F) \right ) \ \le \ A' \ e^{-\mu'
|x-y|}\; ,
\end{equation}
where $\mu^{x,y}_\omega$ is the spectral measure associated to the
pair $x,y$ and $H_\omega$. \label{thm:dynamicallocalization}
\end{thm}

\noindent {\bf Remarks:}

\noindent {\bf 1.}  Recall that for any regular Borel measure
$\mu$, $|\mu|(F) = \sup |\int_F f(E) \mu(dE)|$ where the supremum
ranges over Borel measurable (or even just continuous) functions
$f$ which are point-wise bounded by $1$. Thus
\eq{eq:spectraldecay} implies that
\begin{equation}
\E \left ( \sup_t |<x|f_t(H_\omega) P_{H_\omega \in F} |y> |
\right ) \ \le \ C \  A' e^{-\mu' |x-y|} \;,
\end{equation}
for any uniformly bounded family of Borel functions $\{f_t \}$. In
particular, we may take $f_t(E) = e^{itE}$ for $t \in \R$ to
obtain dynamical localization (\ref{eq:dynamicallocalization}) as
promised.

\noindent {\bf 2:} The requirement that the conditional densities,
$\rho^x_{\omega}$, be uniformly bounded is overly strong.  By the
arguments presented in
ref. \cite{Ai94}, the result extends to potentials for
which there is some $q >0$ such that $ \int (\rho^x_\omega(v))^{1
+ q} dv $ are uniformly bounded.

\noindent {\bf 3:} Since this work extends now the {\it exponential}
dynamical localization to the regime covered by the multiscale analysis,
let us mention that prior results covering this regime
include the proof of localization in terms of {\it power-law} bounds
for the time evolution kernel~\cite{GeDB,DS99}.  (The analysis
there is more general since it applies also to models for which the
fractional moment method has not been developed, e.g., continuum
operators).

\begin{proof_of}{Theorem~\ref{thm:dynamicallocalization}}
It is convenient to derive the result through the analysis of the
finite volume operators obtained by restricting
$H_\omega$ to finite regions,
$\Lambda_n \subset \Z^d$.
It is generally understood that for each $x,y \in \Z^d$ and each
increasing sequence of finite regions $\Lambda_n$ which contain
$\{x,y\}$ and whose union is $\Z^d$, the associated spectral
measures,
$\mu^{x,y}_{\Lambda_n;\omega}$, converge in the vague topology to
$\mu^{x,y}_\omega$. Thus, by the lemma of Fatou, for any
 $F \subset \R$: \  $\E(|\mu^{x,y}_\omega|(F))
\le \liminf_{n\to \infty}  \E(|\mu^{x,y}_{\Lambda_n;\omega}|(F))$.

The upshot is that it suffices to prove the following
statement regarding finite volume operators.

\noindent {\it Under the assumptions
of Theorem~\ref{thm:dynamicallocalization}
there exist $C, \ r > 0$ (which depend only on the
regularity assumptions for $H_\omega$) such that for any finite
region $\Lambda \subset \Z^d$, any $x,y \in \Lambda$, any $F
\subset \R$, and any $s \in (0,1)$}:
\begin{equation}\label{eq:finitevolumebound}
  \E \left ( \left | \mu^{x,y}_{\Lambda;\omega} \right |(F) \right )
  \ \le \ C \ \left [ \sup_{E \in F} \E \left (
  | <x| {1 \over H_{\Lambda,\omega} - E} |y>|^s \right ) \right ]^r \; .
\end{equation}
Following is a summary of the proof of this assertion.

Let us fix a finite region $\Lambda \subset \Z^d$ and a pair of
sites $x,y \in \Lambda$. For simplicity of notation, we will
suppress the region $\Lambda$ and denote the restricted operator
by $H_\omega$ and the associated spectral measure by
$\mu^{x,y}_\omega$

Since $\ell^2(\Lambda)$ is finite dimensional, $\mu^{x,y}_\omega$
is a weighted sum of Dirac measures supported on the
eigenvalues of $H_\omega$.  Integrals with respect to this measure
are discrete sums.  The argument of ref.\cite{Ai94} makes an
essential use of the  following representation of this measure.

\noindent {\it  Let $v  = V_\omega(x)$, and let $\hat v$
be any other value in $ \R$.  Denote
$\widehat \Gamma(E) :=  -1 / <x| {1 \over \widehat
H_\omega - E} |x>$, with  $\hat H_\omega$ the operator
with the potential at $x$ changed to $\hat v$.
Then, }
\begin{equation}\label{eq:spectralintegral}
\mu^{x,y}_\omega(d E) \ = \ - (v - \hat v) \ <x| {1 \over \hat
H_\omega - E} |y> \ \delta(v - \hat v - \hat \Gamma(E)) \ d E \; .
\end{equation}

\noindent In what follows, we will take $\hat v = \hat v_\omega$
to be a random variable independent of $v_\omega$ and identically
distributed. In this case \eq{eq:spectralintegral} holds almost
surely.

A special case of \eq{eq:spectralintegral} is the formula
(which was the basis for the important
``Kotani-argument''\cite{Ko,SiWo})
for the spectral measure at $x$
\begin{equation}\label{eq:spectralintegralxx}
\mu^{x,x}_\omega(d E) \ = \
    \delta(v - \hat v - \hat \Gamma(E)) \ d E \; .
\end{equation}
The above is a probability measure.  Another normalizing
condition is:
\begin{equation}\label{eq:l2bound}
  |v - \hat v|^2 \int |<x| {1 \over \hat H_\omega - E} |y>|^2
  \ \delta(v - \hat v - \hat \Gamma(E)) \ d E \ \le \ 1 \;,
\end{equation}
(which typically holds as equality).

The reason for \eq{eq:l2bound}  is that by the general structure of
the  spectral measures, $\mu^{x,y}_\omega(d E)
=  \Psi_\omega(E) \mu_\omega^{x,x} (dE)$, with $\Psi_\omega(E)$
satisfying $\int |\Psi_\omega(E)|^2
\mu_\omega^{x,x} (d E) \ = \ <y | \, P_\omega \, |y> \ \le \ 1$,
where $P_\omega$ is the projection onto the cyclic
subspace for $H_\omega$ which contains $|x>$.

Let us first present the necessary estimates for the case that $F
\subset \R$ is of finite Lebesgue measure.  Using the bound
\eq{eq:l2bound}, and the H\"older inequality,
\begin{multline}
\E \left ( \left | \mu^{x,y}_\omega \right |(F) \right ) \\ \le \
\left [ \E \left ( |v - \hat v|^{\alpha} \ \int_F |<x| {1 \over
\hat H_\omega - E} |y>|^\alpha \ \delta(v - \hat v - \hat
\Gamma(E)) \ d E \right ) \right ]^{1/(2-\alpha)} \; ,
\end{multline}
where $\alpha \ (\, < 1) $ is a small number to be specified later.
 By a
further application of the H\"older inequality, followed by the
Jensen inequality we obtain
\begin{multline}
 \E \left ( \left | \mu^{x,y}_{\Lambda;\omega} \right |(F) \right
 )^{2-\alpha}
    \ \le \ \left[ 2 \E( |v|^\delta ) \right ]^{\alpha/\delta}
    \\ \times \left [ \E \left ( \int_F |<x| {1 \over
\hat H_\omega - E} |y>|^{s} \ \delta(v - \hat v - \hat \Gamma(E))
\ d E \right )\right]^{\alpha/s} \; ,
\end{multline}
where $\alpha$ is fixed by the equation $ \alpha/s+ \alpha/\delta
= 1$.   Finally we evaluate:
\begin{multline}
  \E \left ( \int_F |<x| {1 \over \hat H_\omega - E} |y>|^{s}
  \ \delta(v - \hat v - \hat \Gamma(E)) \ d E \right ) \\
  \begin{split}
   =& \ \int_F \E \left (
   |<x| {1 \over \hat H_\omega - E} |y>|^{s}
  \rho^x_\omega(\hat v + \hat \Gamma(E))  \right ) \ d E \\
  \le & \ \kappa \ \, \int_{F} \E \left (
   |<x| {1 \over \hat H_\omega - E} |y>|^{s} \right ) \ dE \; ,
  \end{split}
\end{multline}
where $\kappa$ is a uniform upper bound for $\rho^x_\omega$.
These estimates can be
combined to provide a bound of the form \eq{eq:finitevolumebound}
for $F$ a finite interval, which was the case considered in
ref.~\cite{Ai94}.   We shall now improve the argument, to obtain
a statement which covers the case that the localized
spectral regime is unbounded.

Since we do not wish our final estimate to depend on the Lebesgue
measure of $F$, we seek a way of introducing an integrable
weight $h(E)$, so that the final bound
involves the integral of $  h(E) dE$ in place of $dE$. This may be
accomplished with the following inequality:
\begin{equation}\label{eq:gbound}
  \left | \mu^{x,y}_\omega \right | (F) \ \le \
    \left ( <x|\ |g(H)|^{2 p} \ |x> \right )^{1 \over 2 p} \ \left ( \int_F
|g(E)|^{-p'}
    \left | \mu^{x,y}_\omega \right | (d E) \right )^{1 \over p'}
\end{equation}
where $1/p + 1/p' = 1$ and $g$ is any continuous function which is
bounded and bounded away from zero.  To prove \eq{eq:gbound},
write $\left | \mu^{x,y}_\omega \right | (F) \ = \ \int_F g(E) /
g(E) \left | \mu^{x,y}_\omega \right | (dE)$, and apply the
H\"older inequality followed by
\begin{equation}
 \left | \int |g(E)|^p \left | \mu^{x,y}_\omega \right |(dE) \right |
  \ \le \ \left  ( <x| \ |g(H)|^{2p} \ |x> \right )^{1/2} \; .
\end{equation}

It is convenient to choose $g(E)^{2p} = (1 + E^2)$, since $<x|
(1+H_\omega^2)|x> = B + V_\omega(x)^2$ where $B_\omega$ is a
bounded random variable which depends only on the off-diagonal
part of $H_\omega$. Upon taking expectations followed by a further
application of the H\"older inequality this leads to
\begin{multline}
  \E \left ( \left | \mu^{x,y}_\omega \right | (F) \right ) \ \le \
    \left [  \E \left ( \left ( B_\omega + V_\omega(x)^2 \right )^{q \over 2p}
            \right ) \right ]^{1/q} \\
    \times \left [ \E \left ( \left ( \int_F {1 \over (1 + E^2)^{p' \over 2p} }
    \left | \mu^{x,y}_\omega \right | (d E) \right )^{q' \over p'}
    \right )\right ] ^{1/q'}
    \; ,
\end{multline}
where $1/q + 1/q' = 1$.  We estimate the two factors on the right
hand side of this inequality separately.

The first factor can be controlled by choosing $q = p \ \delta$ so
that
\begin{equation}
  \E \left ( \left ( B_\omega + V_\omega(x)^2 \right )^{q \over 2p}
            \right ) \ \le \ \|B_\omega\|_\infty^{\delta / 2} + \E \left
            ( |V_\omega(x)|^\delta
            \right ) \; .
\end{equation}
The exponents $p$, $p'$, $q$, $q'$ are all specified once we
choose $p> 1/\delta$. Specifically, $q = \delta p$, $q' = p (p -
1/\delta)^{-1}$, and $p' = p (p - 1)^{-1}$.  Note that $p' < q'$.

To estimate the second factor, we note that $|\mu^{x,y}_\omega|$
is a sub-probability measure and $ q'/p' >1$, so by the Jensen
inequality,
\begin{equation}
  \E \left ( \left ( \int_F {1 \over (1 + E^2)^{p' \over 2p} }
    \left | \mu^{x,y}_\omega \right | (d E) \right )^{q' \over p'}
    \right ) \ \le \ \E \left ( \int_F {1 \over (1 + E^2)^{q' \over 2p} }
    \left | \mu^{x,y}_\omega \right | (d E) \right ) \;.
\end{equation}
Estimating the right hand side with the argument outlined above
for $F$ with finite Lebesgue measure, we find that
\begin{multline}
  \E \left ( \int_F {1 \over (1 + E^2)^{q' \over 2p} }
    \left | \mu^{x,y}_\omega \right | (d E) \right ) \ \le \
    \left[ 2 \E( |v|^\delta ) \right ]^{\alpha/\delta} \\
    \times \left [ \kappa \   \int_F
    \E \left ( |<x| {1 \over \hat H_\omega - E} |y>|^{s} \right )
      \, {dE \over (1+E^2)^{q'/2p} } \right ]^{\alpha/s} \; ,
\end{multline}
which is uniformly bounded provided we choose $p$ such that $q'/p
> 1$.  This is possible since $q'/p = (p - 1/\delta )^{-1}$ which
can be made as large as we like.

Thus, for any finite volume $\E \left ( \left |
\mu^{x,y}_{\Lambda;\omega} \right | (F) \right )$ can be bounded
by a constant multiple of $\sup_{E \in F} \E \left ( |<x| {1 \over
\hat H_{\Lambda;\omega} - E} |y>|^{s} \right )$ raised to a
certain power. Which multiple and which power depend only on the
$\delta$-moments of the potential and the uniform bound on the
conditional distributions $\rho^x_\omega$. By the vague
convergence argument outlined at the start of the proof, this
proves the theorem.
\end{proof_of}

\maappendix{A fractional moment bound}
\label{sect:moment}

The regularity conditions $R_1(\tau)$ and $R_2(s)$ have been used
to give {\em a priori} estimates of certain fractional moments.
Such fractional moment bounds are properties of the general class
of operators with diagonal disorder. Hence, throughout this
appendix, we consider random operators $H_\omega$ on $\ell^2(\T)$
of the form
\begin{equation}
    H_\omega \ = \ T_0 + \lambda V_\omega \; , \label{eq:arbitrary}
\end{equation}
where $T_0$ is an arbitrary bounded self--adjoint operator and
$V_{\omega}$ is a random potential  for which $V_\omega(x)$ are
independent random variables ($\T$ is any countable set).

\begin{lem}
\label{lem:fracmom} Let $H_{\omega}$ be a random operator given by
\eq{eq:arbitrary} such that for each $x$ the probability
distribution of the potential $V_\omega(x)$ satisfies $R_1(\tau)$
for some fixed $\tau>0$ with constants uniform in $x$. Then there
exists $\kappa_\tau < \infty$ such that for any finite subset
$\Lambda$ of $\T$, any $x, y \in \Lambda$, any $z \in \mathbb C$,
and any $s \in (0, \tau)$
\begin{equation}
\label{eq:fracmom} \E \left ( \left . |<x|
\frac{1}{H_{\Lambda;\omega} - z} |y>|^s \right | \{V(u)\}_{u \in
\Lambda \backslash \{x,y\}} \right) \ \leq \ {\tau \over \tau - s}
\
 { (4 \kappa_{\tau}) \over \lambda^s}^{s/\tau} \; .
\end{equation}
\end{lem}
\begin{proof}
Let us first consider $z=E \in \R$. For such energies
\eq{eq:fracmom} is a consequence of a Wegner type estimate on the
2-dimensional subspace spanned by $|x>, |y>$. The key is to
determine the correct expression for the dependence of $<x|
\frac{1}{H_{\Lambda;\omega} - E} |y>$ on $V_\omega(x)$ and
$V_\omega(y)$. Such an expression is given by the `Krein formula':
\begin{equation}
 <x| \frac{1}{H_{\Lambda;\omega} - E} |y> \ = \
    <1| \left ( [A]^{-1} +  \lambda \begin{bmatrix}
        V_\omega(x)&0\\0 & V_\omega(y)
        \end{bmatrix}
     \right )^{-1} |2> \; ,
\label{eq:Krein}
\end{equation}
where $[A]$ is a $2\times2$ matrix whose entries do not depend on
$V_\omega(x)$ or $V_\omega(y)$. In fact,
\begin{equation}
    [A] \ = \ \begin{bmatrix}
      <x|\frac{1}{\widehat H_{\Lambda;\omega} - E} |x> &
      <x|\frac{1}{\widehat H_{\Lambda;\omega} - E} |y> \\
      <y|\frac{1}{\widehat H_{\Lambda;\omega} - E} |x> &
      <y|\frac{1}{\widehat H_{\Lambda;\omega} - E} |y>
    \end{bmatrix} \; ,
\end{equation}
where $\widehat H_{\Lambda;\omega}$ denotes the operator obtained
from $H_{\Lambda;\omega}$ by setting $V_\omega(x)$ and
$V_\omega(y)$ equal to zero.

The regularity condition $R_1(\tau)$ implies a Wegner type
estimate: 
\begin{equation}
\P \left ( \left \| \left ( [A]^{-1} + \lambda
        \begin{bmatrix}
          V_\omega(x)&0 \\
          0 & V_\omega(y)
        \end{bmatrix}
     \right )^{-1} \right \| > t
        \left . \phantom{a \over b} \right | \; \{V_\omega(u)\}_{u \neq
x,y} \right )
     \ \leq \ {4 \kappa_{\tau} \over (\lambda t)^{\tau}} \; ,
\label{eq:Wegner}
\end{equation}
where $\kappa_\tau$ is any finite number such that for every $v
\in \T$, $a \in \R$, and $\epsilon > 0$
\begin{equation}
 \P \left ( V_\omega(v) \in
(a-\epsilon,a+\epsilon)  \right ) \ \le \ \kappa_\tau
\epsilon^{\tau} \; .
\end{equation}
The desired bound (\ref{eq:fracmom}) follows easily from
\eq{eq:Wegner}. (The factor, $4$, on the right hand side of
(\ref{eq:Wegner}) arises as the square of the ``volume'' of the
region $\{x,y\}$. In the case $x = y$, we could replace this
factor by $1$.)

Although the Krein formula (\ref{eq:Krein}) is true when $E$ is
replaced by any $z \in \C$, the resulting matrix $[A]$ may not be
normal if $z \not \in \R$.  (The resolvent, $1 \over H - z$, {\em
is} normal. However, given an orthogonal projection, $P$, the
operator $P {1 \over H - E} P$ may not be normal!) Yet, the
Wegner-like estimate (\ref{eq:Wegner}) holds only when $[A]$ is a
normal matrix.  At first, this seems to be an obstacle to the
extension of (\ref{eq:fracmom}) to all values of $z$. However,
once the inequality is known for real values of $z$, it follows
for all $z \in \mathbb C$ from analytic properties of the
resolvent. Specifically, the function
\begin{equation}
\phi(z) \ = \ |<x| \frac{1}{H_{\Lambda;\omega} - z} |y>|^s
\end{equation}
is {\em sub-harmonic} in the upper and lower half planes and
decays as $z \rightarrow \infty$.  Hence, $\phi(z)$ is dominated
by the convolution of its boundary values with a Poisson kernel:
\begin{equation}
\phi( E + i \eta) \ \leq \ \int \phi(\widetilde E) {|\eta| \over
(E - \widetilde E)^2 + \eta^2}{d \widetilde E \over \pi} \; .
\end{equation}
By Fubini's theorem and \eq{eq:fracmom} for $\widetilde E \in \R$,
(\ref{eq:fracmom}) is seen to hold for all $z \in \mathbb C$.
\end{proof}

The ``all for one'' principle mentioned previously is actually a
simple consequence of Lemma~\ref{lem:fracmom}.
\begin{lem}
Let $H_{\omega}$ be a random operator as described in
Lemma~\ref{lem:fracmom}, and suppose that there is a distance
function $\dist$ on $\T$ such that for some $s < \tau$ and some $z
\in \C$
\begin{equation}
\E\left( |<x| {1 \over  H_{\omega}-z } |y> |^s \right) \ \ \le \ \
A(s) \  e^{-\mu(s) \, \dist(x,y) } \; , \label{eq:allforone}
\end{equation}
for every $x,y \in \T$. Then, in fact, (\ref{eq:allforone}) holds,
with modified constants $A(r)$ and $\mu(r)$, when $s$ is replaced
by any $r < \tau$. \label{lem:allforone}
\end{lem}
\begin{proof}
Note that given $r,s>0$ with $r < s < \tau$
\begin{multline}
\E \left(|<x| {1 \over  H_{\Lambda;\omega}-E } |y> |^r \right)^{s
\over r} \ \le \ \E \left(
        |<x|{1 \over H_{\Lambda;\omega}-E } |y> |^s
        \right) \\ \le \ \E \left(
        |<x| {1 \over  H_{\Lambda;\omega}-E } |y> |^r
        \right)^{t-s \over t- r}
        \E \left(
        |<x|{1 \over H_{\Lambda;\omega}-E } |y> |^t
        \right)^{s- r \over t - r} \\ \le \
        \left (
        {(4 \kappa_{\tau}) \over \lambda^t}^{t/\tau}
        \right )^{s- r \over t - r}
        \E \left(
        |<x| {1 \over  H_{\Lambda;\omega}-E } |y> |^r
        \right)^{t- s \over t- r} \; ,
\end{multline}
where $t$ is any number with $s < t < \tau$.
\end{proof}

\maappendix{Decoupling inequalities} \label{sect:decoupling}

\masubappendix{Decoupling inequalities for Green Functions}

The condition $R_2(s)$ plays a crucial role in several of the
arguments presented in this paper.  It has been used to bound
expectations of products of Green functions in terms of products
of expectations.  In this section we demonstrate the validity of
the necessary bounds. The main result is the following:

\begin{lem}
Let $H_{\omega}$ be a random operator given by \eq{eq:arbitrary},
with an $s$ regular distribution of the potential $V_\omega(x)$.
Then 
\begin{enumerate}
\item For any $\Omega_1, \Omega_2 \subset \T$, any $x,y \in
\Omega_1$, and any $u,v \in \Omega_2$,
\begin{equation}
\E \left ( |G_{\Omega_1}(x,y;z)|^s |G_{\Omega_2}(u,v;z)|^s \right
) \ \le \ {\widetilde C_s \over \lambda^s} \ \E \left (
|G_{\Omega_1}(x,y;z)|^s \right ) \; .
\end{equation}
\item  For any $\Omega_1 \cap \Omega_2 = \emptyset$,
$x,u \in \Omega_1$, $v,y \in \Omega_2$, and $\Omega_3 \subset
\Gamma$,
\begin{multline}
\E \left ( |G_{\Omega_1}(x,u;z)|^s |G_{\Omega_3}(u,v;z)|^s
|G_{\Omega_2}(v,y;z)|^s \right ) \\ \le \ {\widetilde C_s \over
\lambda^s} \ \E \left ( |G_{\Omega_1}(x,u;z)|^s \right ) \E \left
( |G_{\Omega_2}(v,y;z)|^s \right ) \; .
\end{multline}
\end{enumerate}
\label{lem:decouplinginequalities}
\end{lem}

Lemma \ref{lem:decouplinginequalities} is a consequence of the
conditional expectation bound (\ref{eq:fracmom}), the Krein
formula (\ref{eq:Krein}), and the following:
\begin{lem}
Let $V_1,V_2$ be independent real valued random variables which
satisfy $R_2(s)$ for some $s > 0$. Then there exists $D^{(2)}_s>
0$ such that
\begin{equation}
\E \left ( |F(V_1,V_2)|^s |F(V_1,V_2)|^s \right ) \ \le \
D^{(2)}_s \ \E \left ( |F(V_1, V_2)|^s \right ) \ \E \left (
|G(V_1, V_2)|^s \right ) \ ,
\end{equation}
where $F$ and $G$ are arbitrary functions of the form
\begin{eqnarray}
F(V_1,V_2) &=& {1 \over L_1(V_1,V_2)} \\ G(V_1,V_2) &=&
{L_2(V_1,V_2) \over  L_3(V_1,V_2) } \; ,
\end{eqnarray}
with $\{L_i\}$ functions which are linear in each variable
separately.  In fact, we may take $D^{(2)}_s = D_{s;1} D_{s;2}$ ,
where, for $j=1,2$, $D_{s;j}$ is the decoupling constant for
$V_j$.
\end{lem}

\begin{proof}
Let $f(V)$ and $g(V)$ be two functions of the appropriate form for
the decoupling lemma. Then, with $j=1,2$
\begin{equation}
\E \left ( |f(V_j)|^s |g(V_j)|^s \right ) \ \le \ D_{s;1} \ \E
\left ( |f(\widetilde V_j) |^s |g(V_j)|^s \right ) \; ,
\end{equation}
where $\widetilde V_j$ indicates an independent variable
distributed identically to $V_j$.

Now, if $F$ and $G$ are functions of $2$ variables of the given
form, then at fixed values of $V_2$, they satisfy the $1$ variable
decoupling lemma, so
\begin{equation}
\E \left ( |F(V_1,V_2)|^s |G(V_1,V_2)|^s \right ) \ \le \ D_{s;1}
\ \E \left (|F(\widetilde V_1, V_2)|^s |G(V_1, V_2)|^s \right ) \;
.
\end{equation}
For fixed values of $\widetilde V_1$ and $V_1$, $F(\widetilde V_1,
V_2)$ and $G(V_1, V_2)$ (as functions of $V_2$) are again of the
correct form to apply the $1$ variable decoupling lemma.  Thus,
\begin{multline}
\E \left ( |F(V_1,V_2)|^s |G(V_1,V_2)|^s \right ) \ \le \ D_{s;1}
D_{s;2} \ \E \left (|F(\widetilde V_1, \widetilde V_2)|^s |G(V_1,
V_2)|^s \right ) \\ = \ D_{s;1} D_{s;2} \ \E \left (|F( V_1,
V_2)|^s  \right ) \ \E \left ( |G(V_1, V_2)|^s \right ) \; .
\end{multline}
\end{proof}

\masubappendix{A condition for the validity of $R_2(s)$}

Decoupling lemmas have been discussed already in
references~\cite{AM,Ai94,AG}.
Though these  contain results similar to those
required here, they do not provide the exact condition
used in this work. Hence, we briefly present an
elementary condition under which $R_2(s)$ is satisfied.
The following discussion is by
no means exhaustive. Rather, we simply wish to show that the
condition $R_2(s)$ is not devoid of meaningful examples.

\begin{lem}
Let $\rho$ be a measure with bounded support which satisfies
$R_1(\tau)$. Then for any $s < {\tau \over 4}$, $\rho$ satisfies
$R_2(s)$.
\end{lem}
\begin{proof}
For each $s> 0$, we define
\begin{eqnarray}
\phi_s(z) &=& \int {1 \over |V - z|^s} \rho(dV) \; , \\
\psi_s(z,w) &=& \int {|V-z|^s \over |V-w|^s} \rho(dV) \; , \\
\gamma_s(z,w,\zeta), &=& \int {|V-z|^s \over |V-w|^s |V -
\zeta|^s} \rho(dV) \; .
\end{eqnarray}
Property $R_2(s)$ amounts to the statement that
\begin{equation}
\sup_{z,w, \zeta \in \mathbb C} { \gamma_s(z,w,\zeta) \over
\phi_s(\zeta) \psi_s(z,w)} \ < \ \infty \; . \label{eq:r2true}
\end{equation}

In fact, if we let
\begin{eqnarray}
F_s(z) \ = \ {\sqrt{\phi_{2s}(z)} \over \phi_s(z)} \; ,
\\ G_s(z,w) \ = \ { \sqrt{
\psi_{2s}(z,w)} \over \psi_s(z,w)} \; ,
\end{eqnarray}
then by the Cauchy-Schwartz inequality, it suffices to show that
$F_s$ and $G_s$ are uniformly bounded.  However this is elementary
since $F_s$ and $G_s$ are continuous functions which are easily
shown to have finite limits at infinity.
\end{proof}



\noindent {\large \bf Acknowledgments\/} \\
This work  was supported in part by the NSF Grant PHY-9971149
(MA).  Jeff Schenker thanks the NSF for
financial support under a Graduate Research Fellowship,
and Dirk Hundertmark thanks the Deutsche Forschungsgemeinschaft
for financial support under grant Hu 773/1-1.

\bigskip

 \addcontentsline{toc}{section}{References}

\end{document}